\documentclass[twocolumn,preprintnumbers,prl,aps]{revtex4-1}
\usepackage{graphicx,amsmath,bbm,color,float}
\usepackage[breaklinks=true]{hyperref}
\graphicspath{ {fig/} }

\allowdisplaybreaks[4]

\definecolor{darkblue}{rgb}{0.0,0.0,0.5}
\newcommand{\txblu}{\textcolor{darkblue}}
\newcommand{\Graph}[2]{\vcenter{\hbox{\includegraphics[scale=#1]{#2}}}}
\newcommand{\ff}{\bar{\mathcal{F}}}
\newcommand{\figgraph}[3]{\overset{(#3-2\epsilon)}{\Graph{#1}{#2}}}

\begin{document}

\preprint{MSUHEP-19-005}

\title{\boldmath
Quark and gluon form factors in four loop QCD: the $N_f^2$ and
$N_{q\gamma} N_f$ contributions
\unboldmath}

\author{Andreas von Manteuffel and Robert M. Schabinger} 

\affiliation{
Department of Physics and Astronomy, Michigan State University, East Lansing, Michigan 48824, USA}

\begin{abstract}
\noindent
We calculate the four-loop massless QCD corrections with two closed quark lines
to quark and gluon form factors.
The results for the gluon form factor and the singlet part of the quark form
factor are given for the first time.
From our analytic expressions for the form factors, we determine the corresponding cusp anomalous dimensions.
The relevant Feynman integrals are obtained with novel integral reduction techniques and direct integration methods.
\end{abstract}

\maketitle
\section{Introduction}

\begin{figure*}
\centering
\begin{align*}
&\quad\Graph{.175}{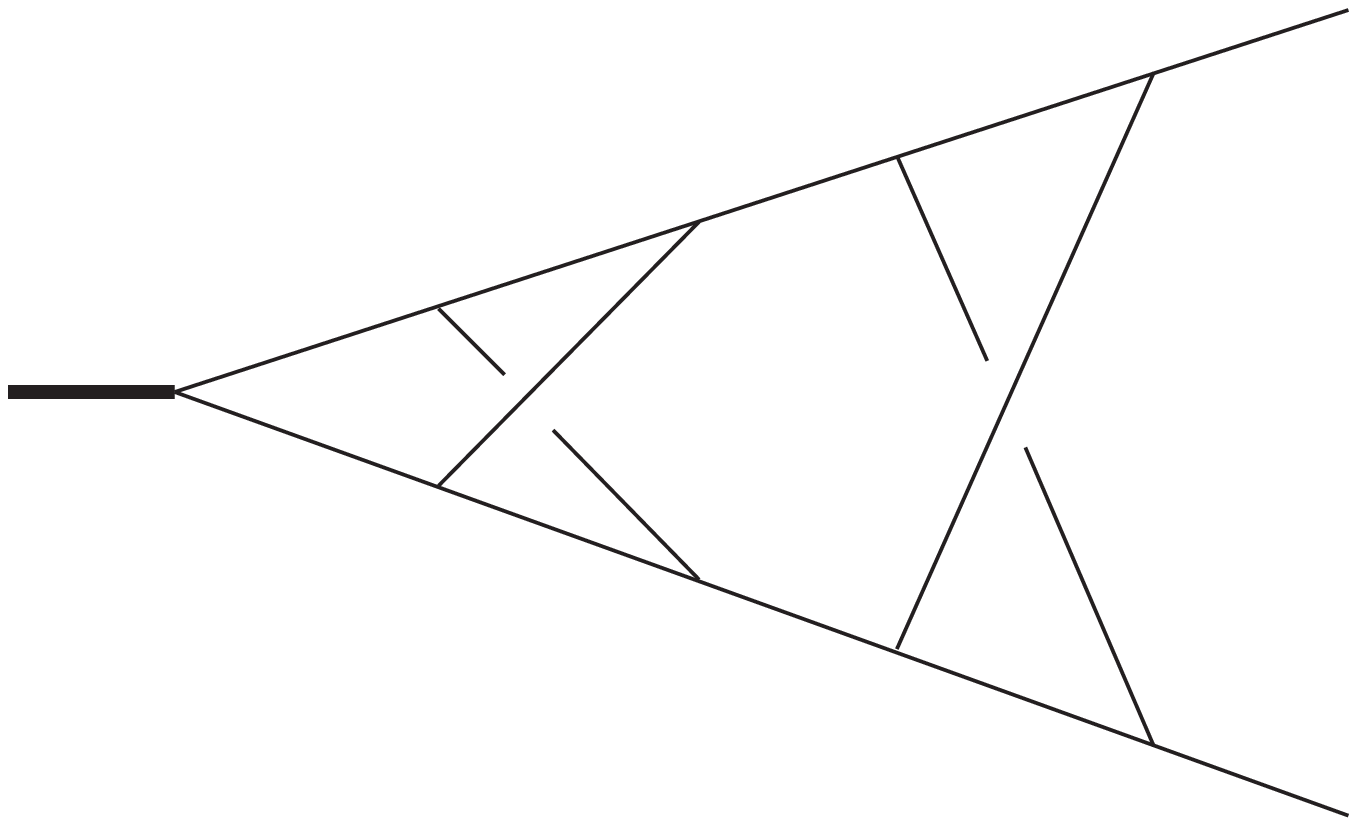} \quad \Graph{.175}{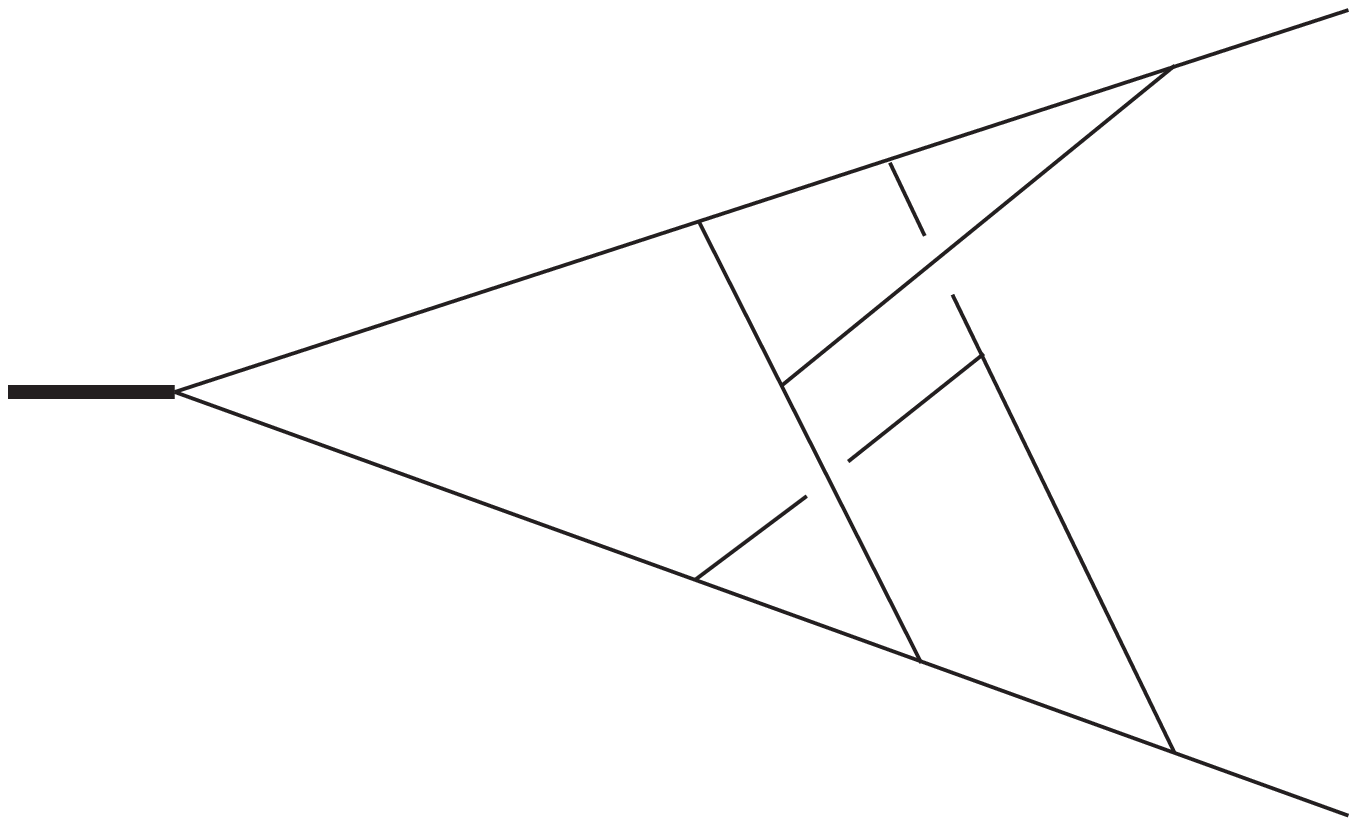} \quad \Graph{.175}{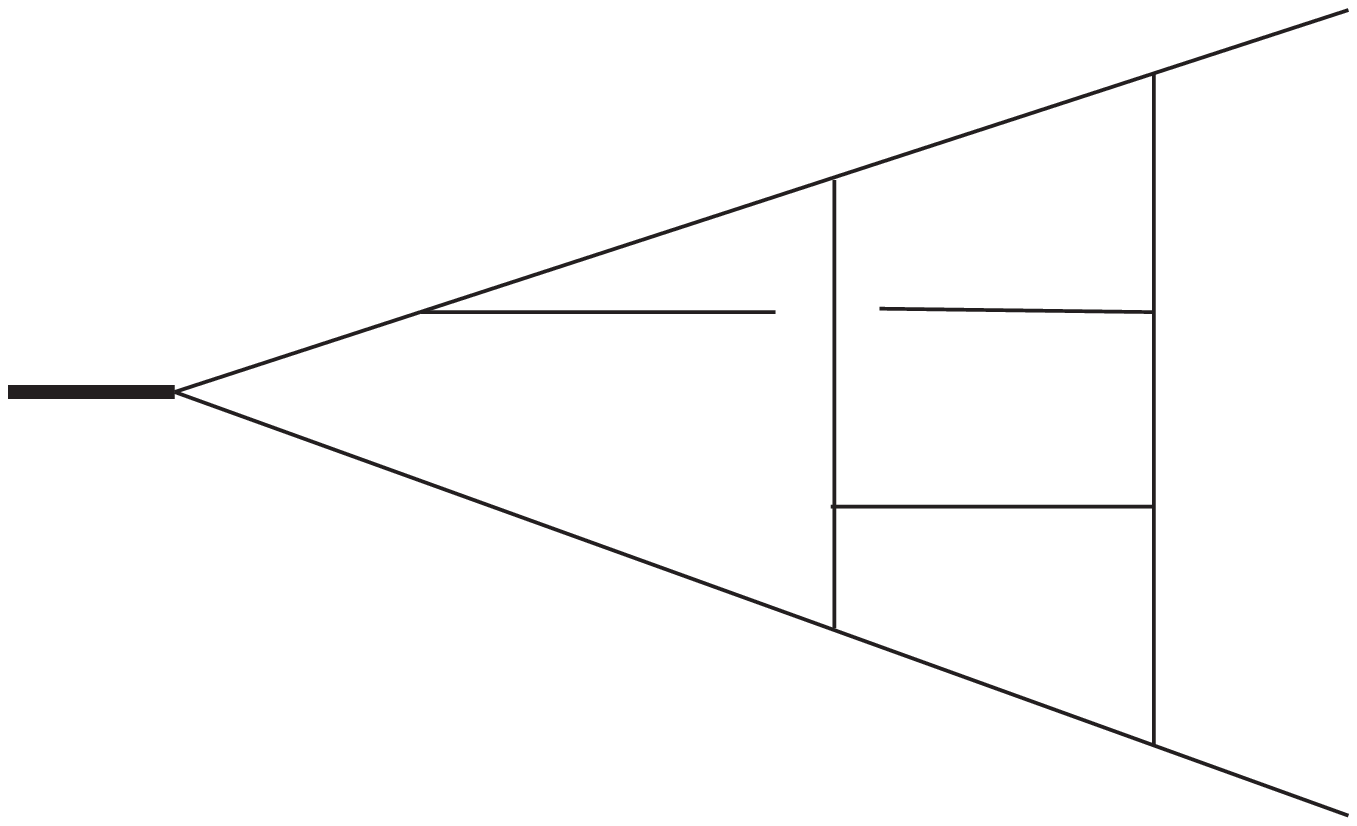} \quad \Graph{.175}{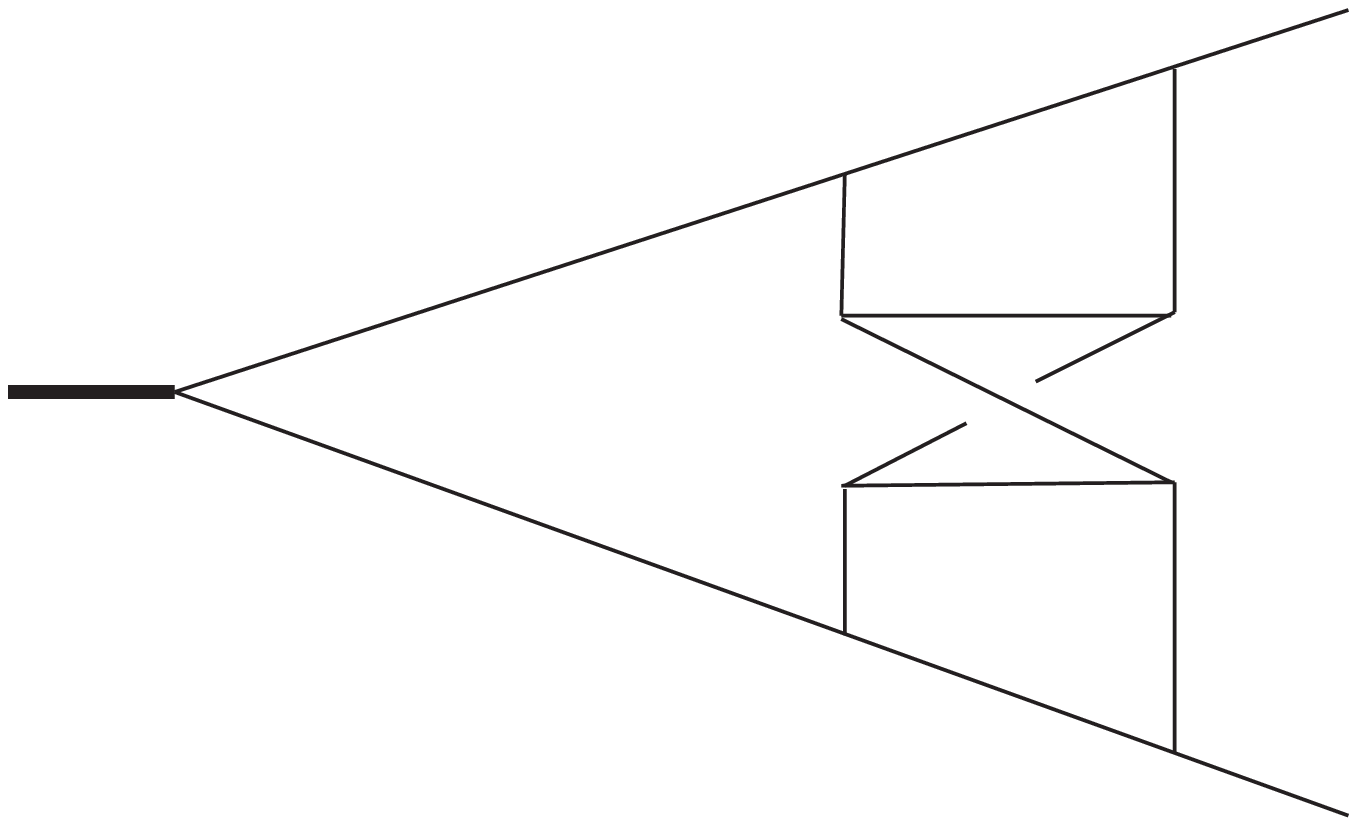}\quad \Graph{.175}{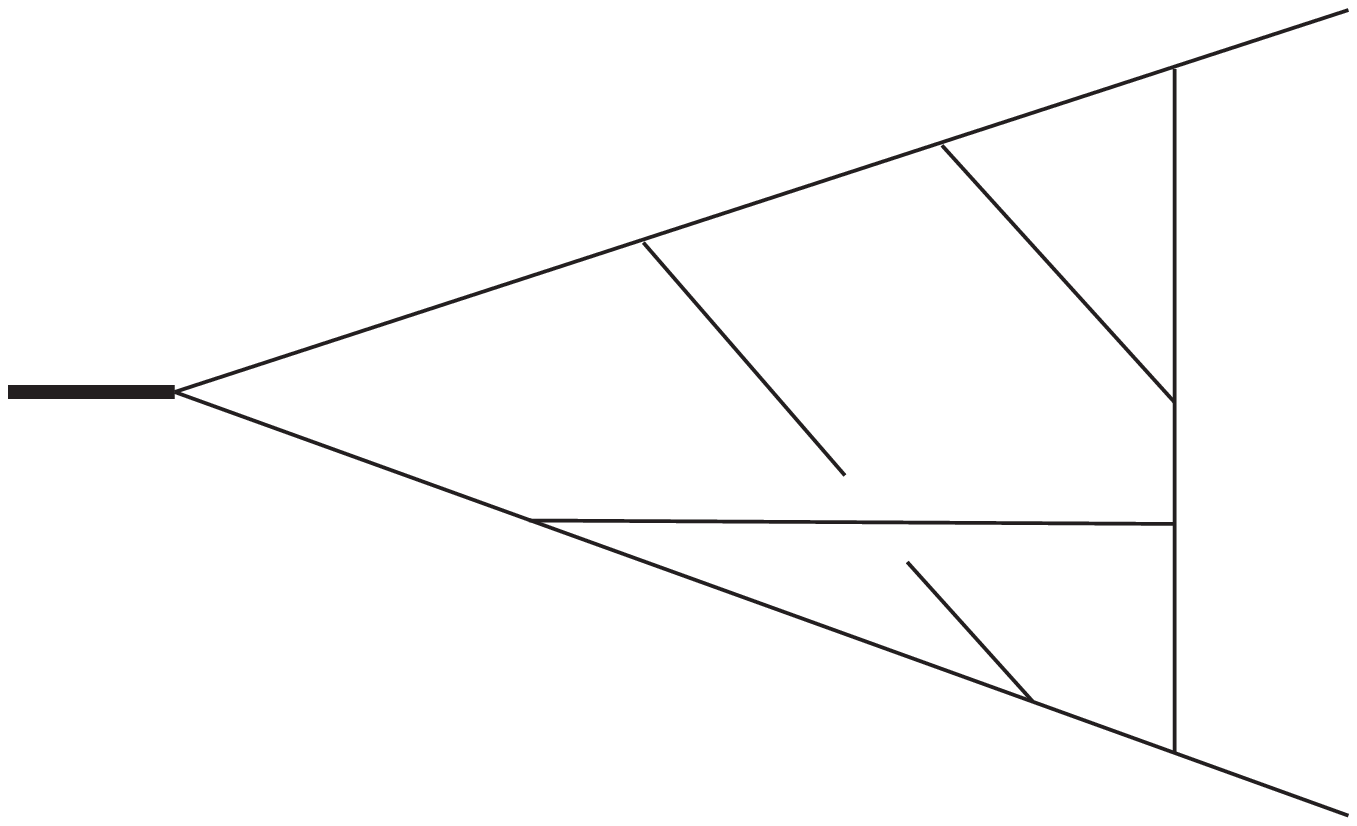} \quad \Graph{.175}{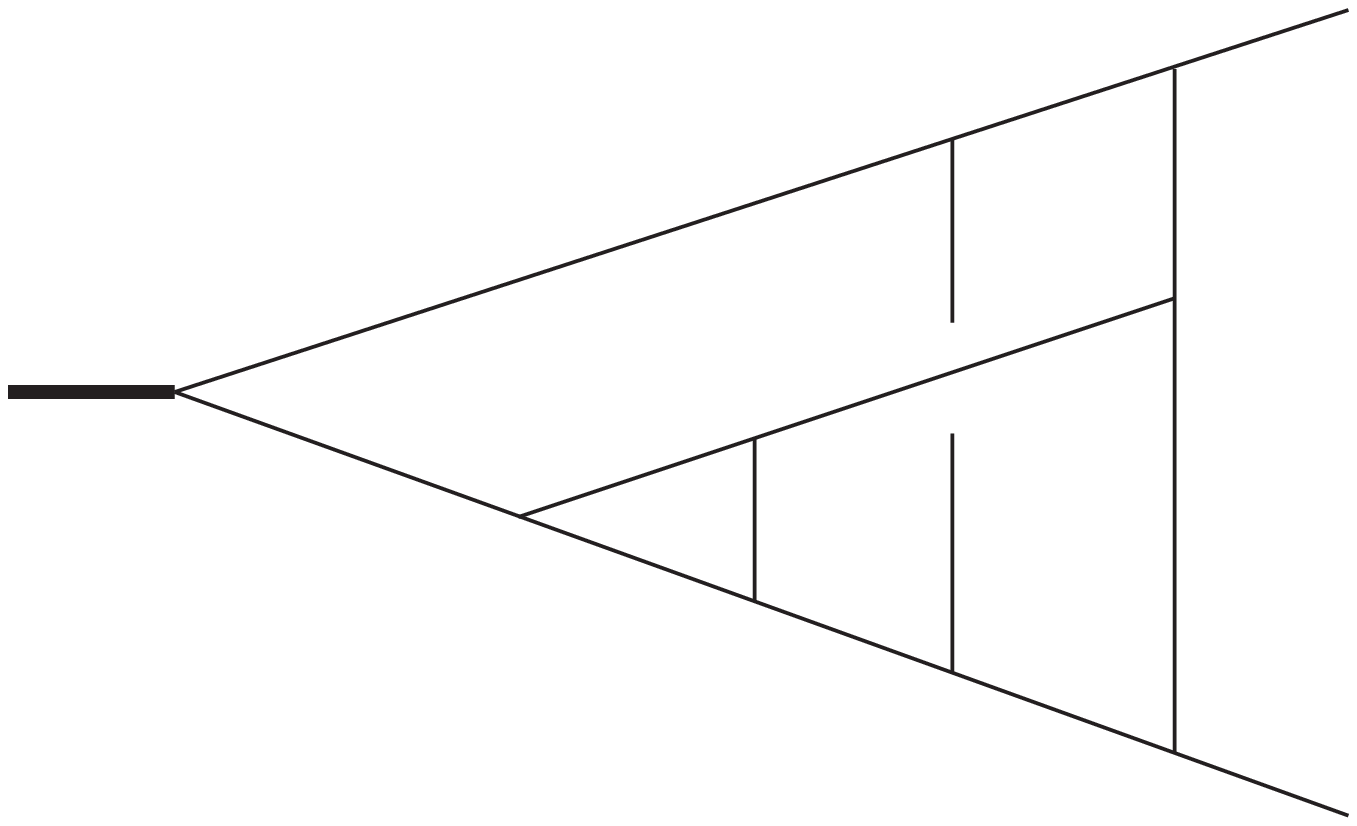}
\\
&\quad\Graph{.175}{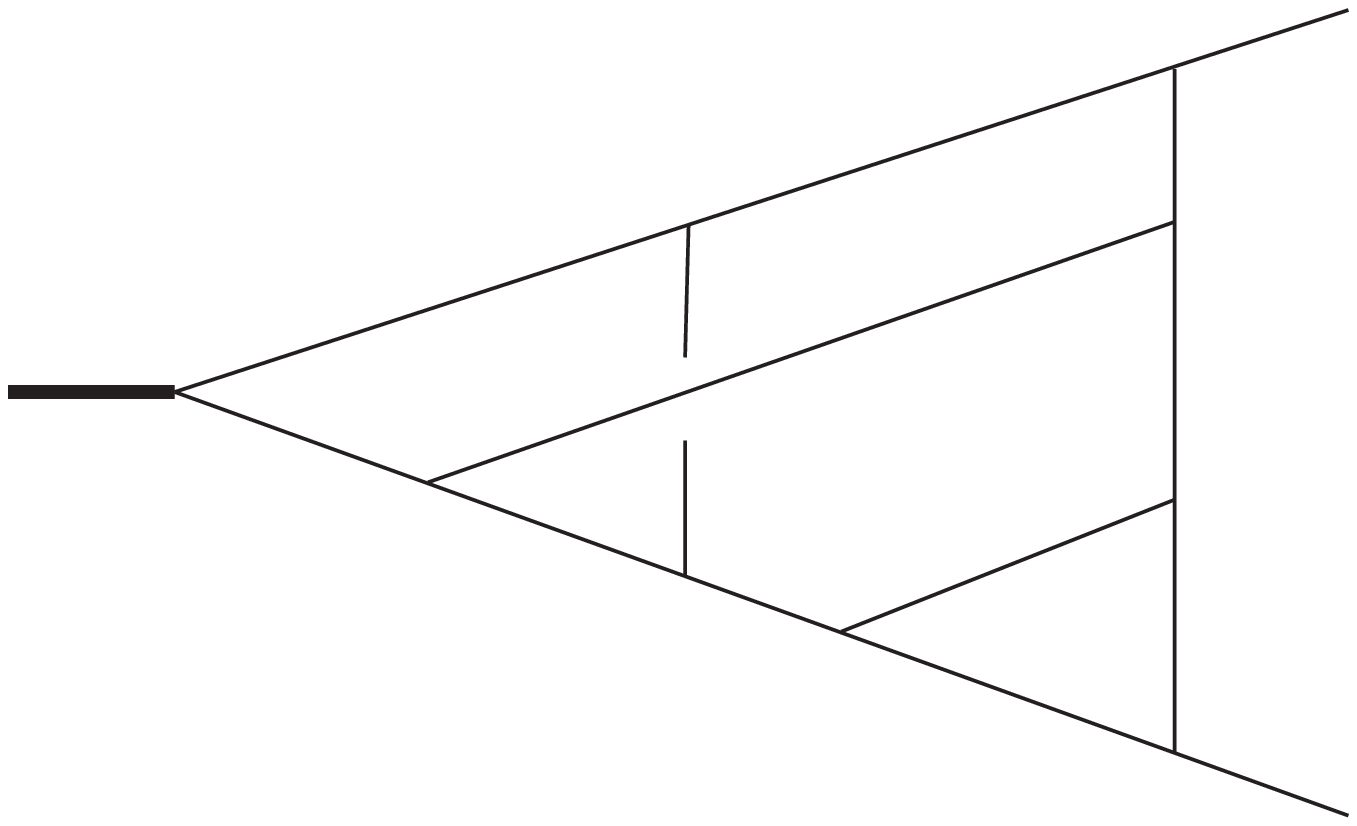} \quad \Graph{.175}{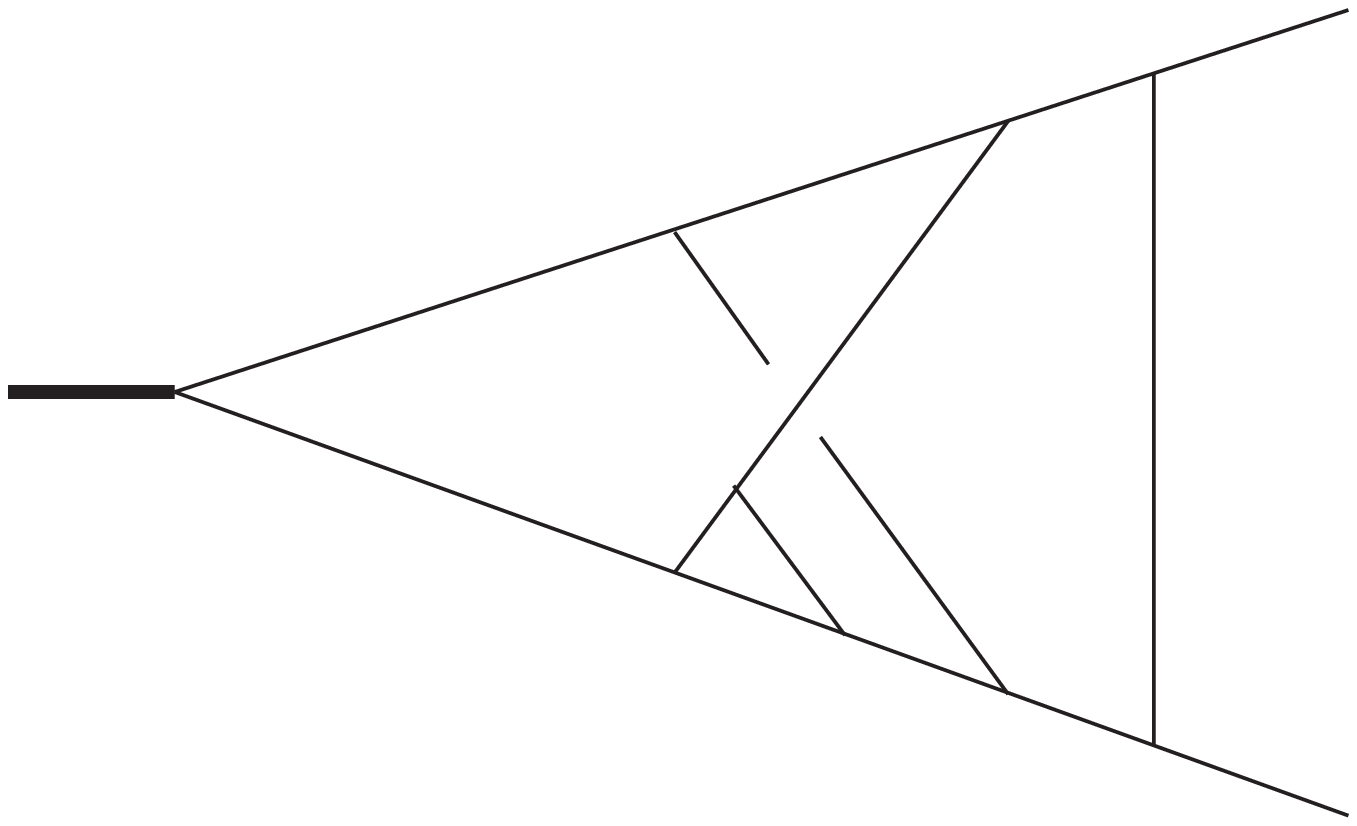} \quad \Graph{.175}{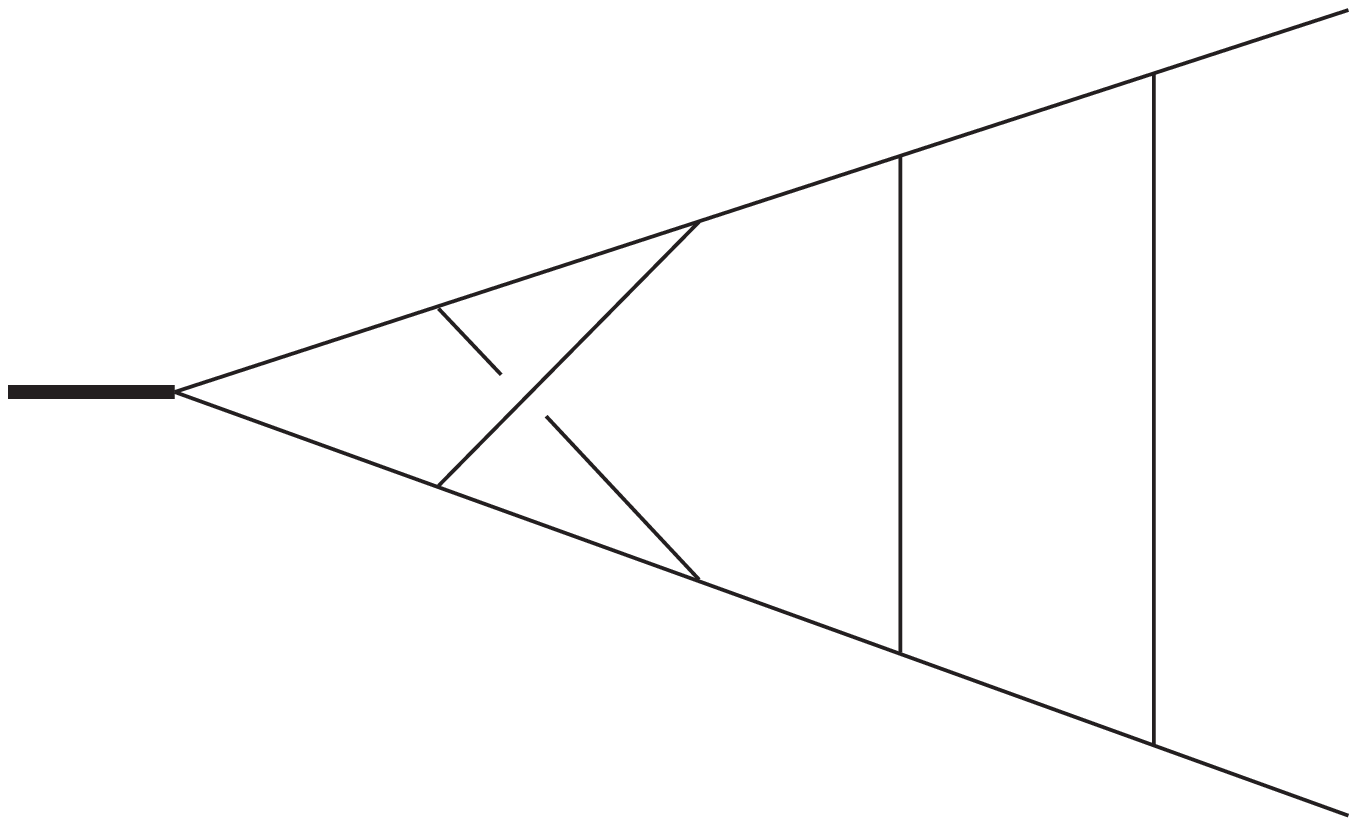} \quad \Graph{.175}{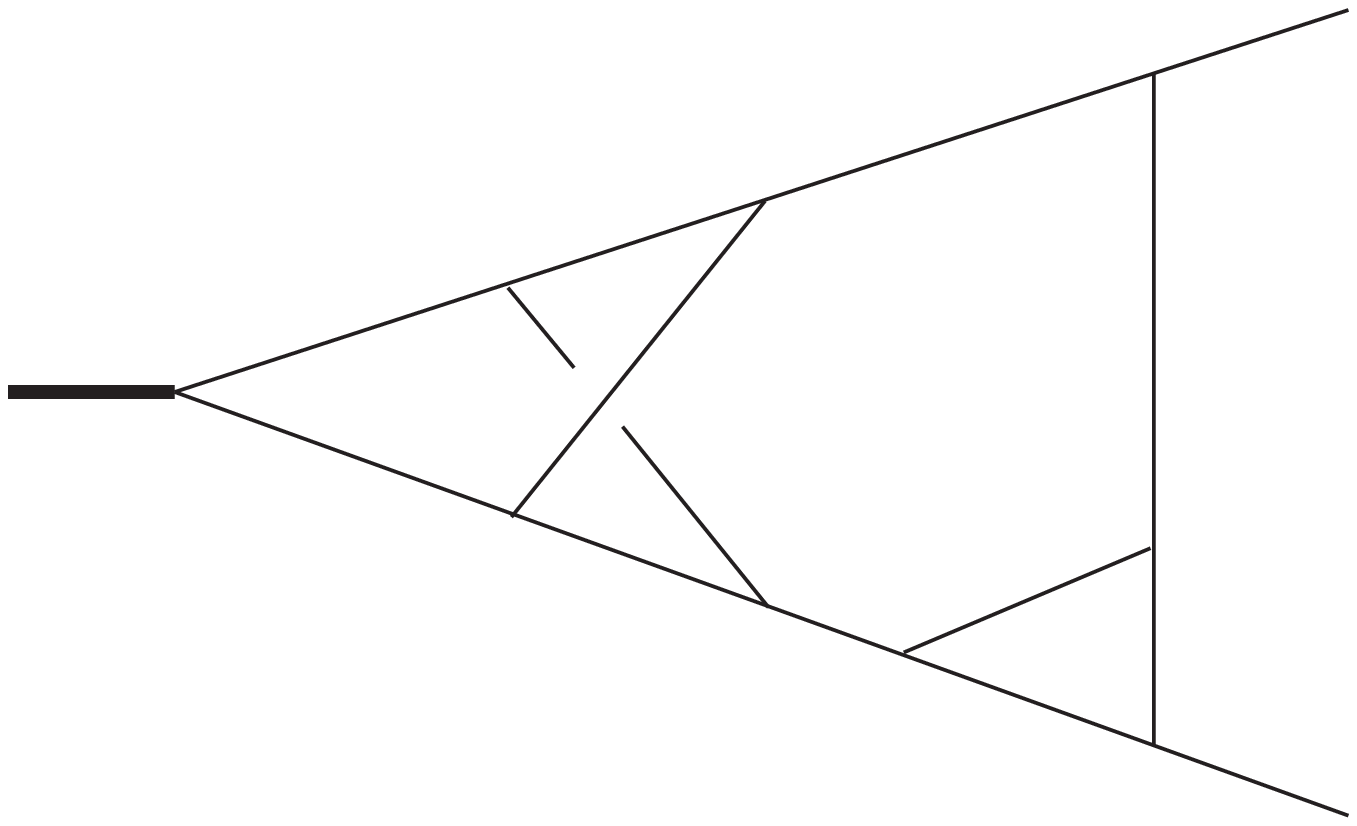}\quad \Graph{.175}{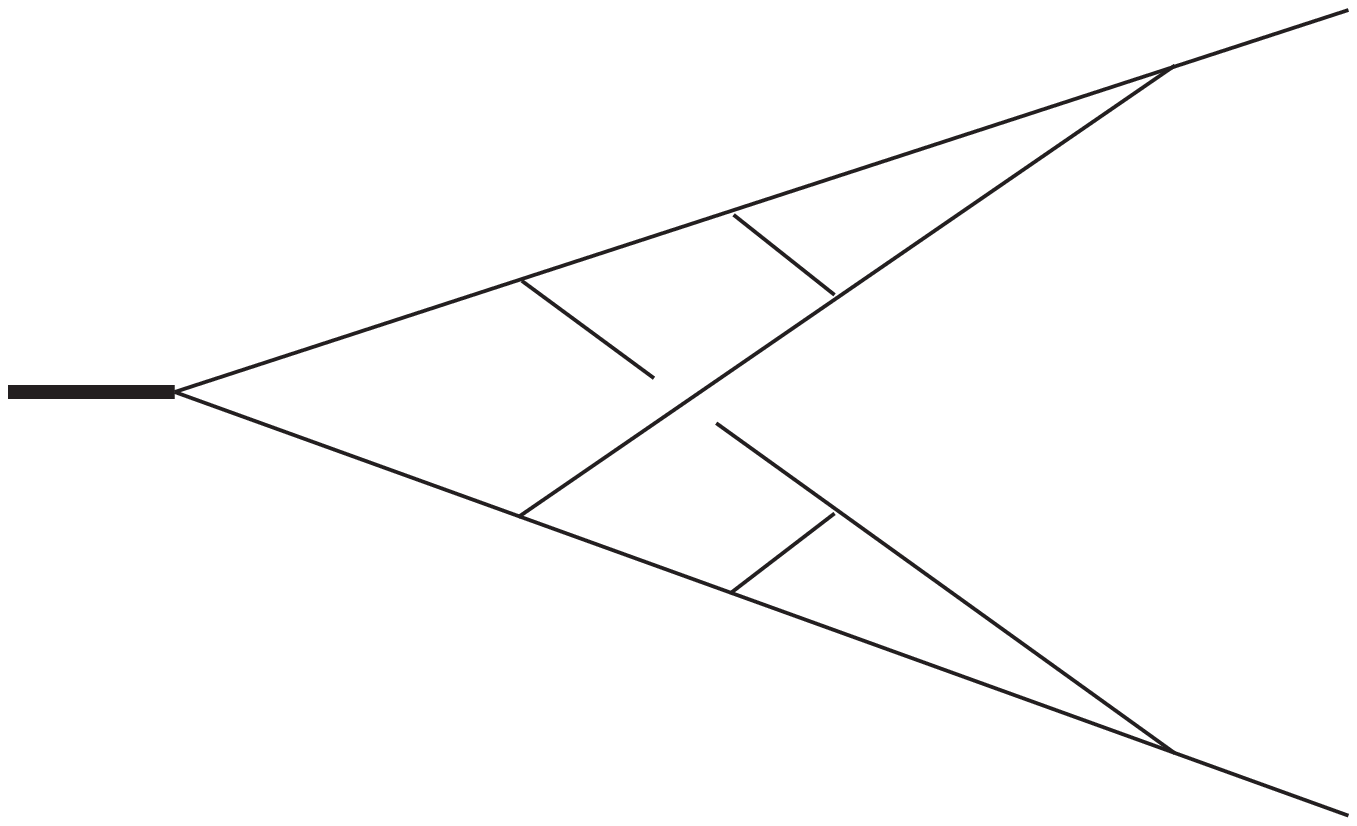} \quad \Graph{.175}{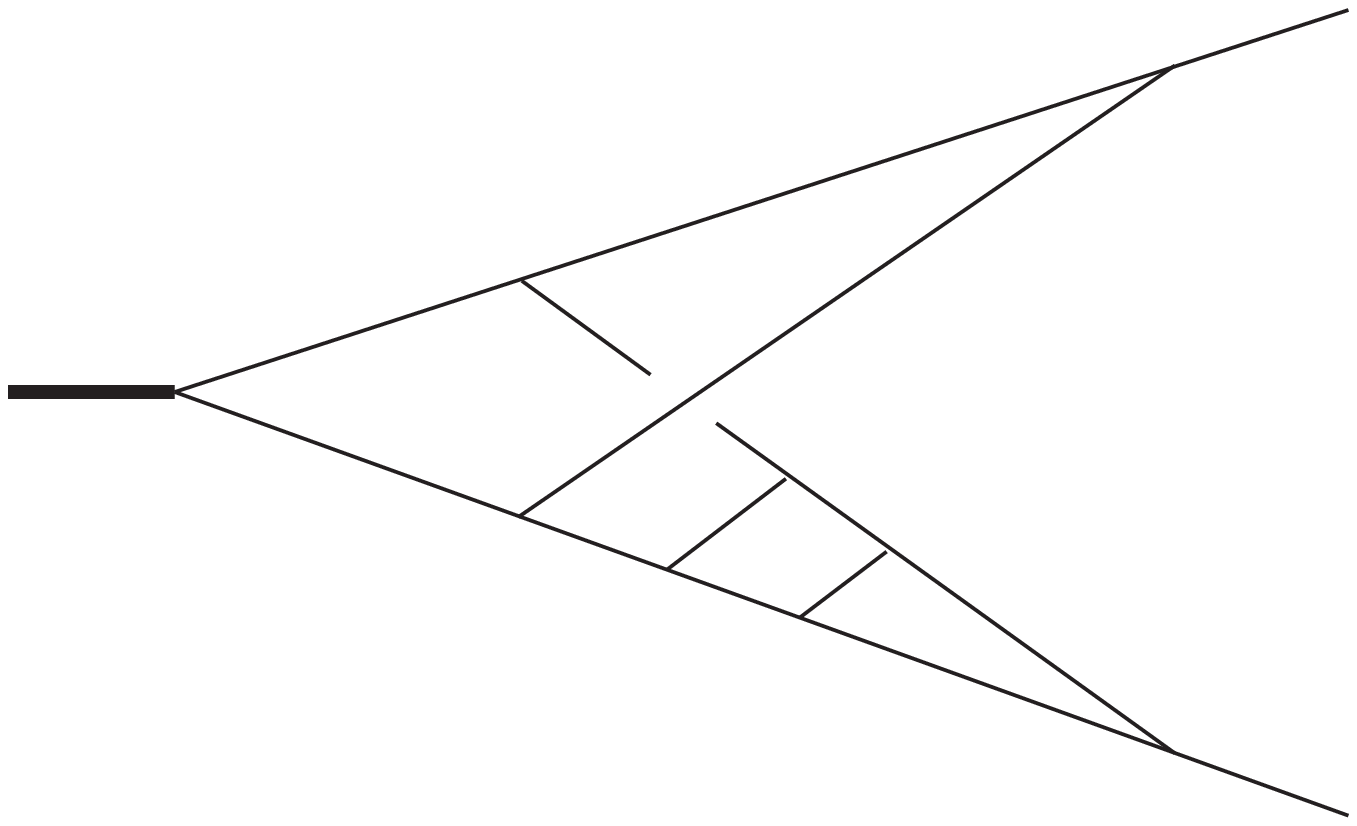}
\\
&\quad\Graph{.175}{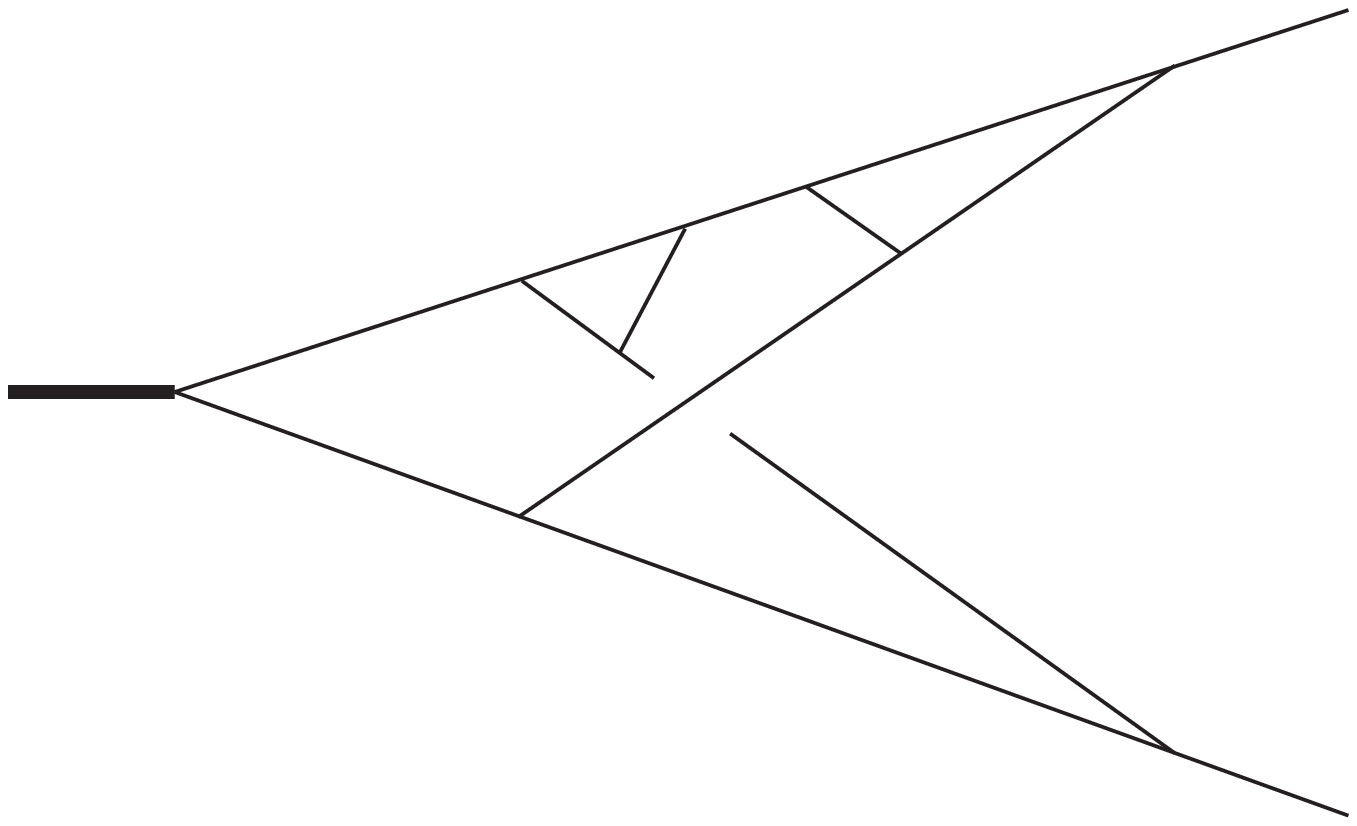} \quad \Graph{.175}{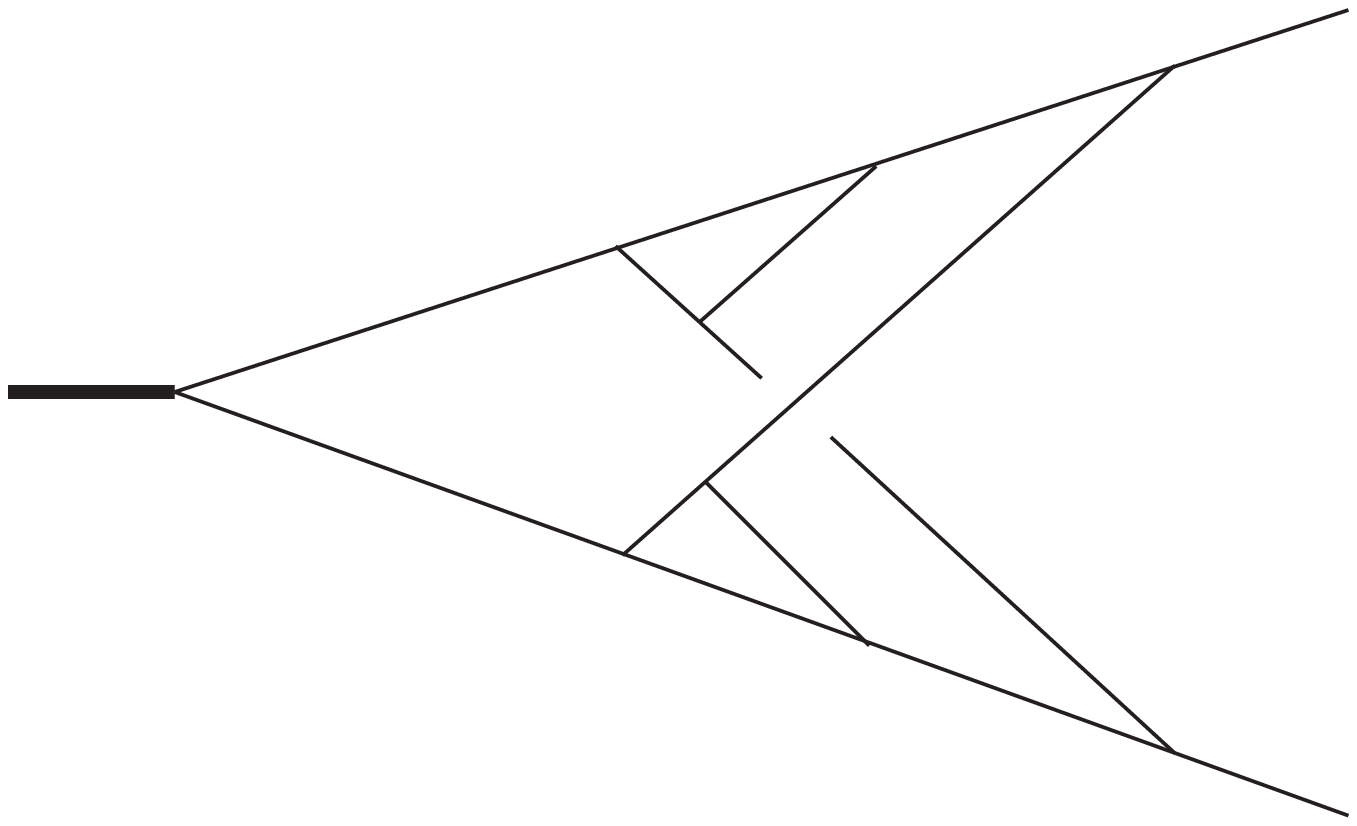} \quad \Graph{.175}{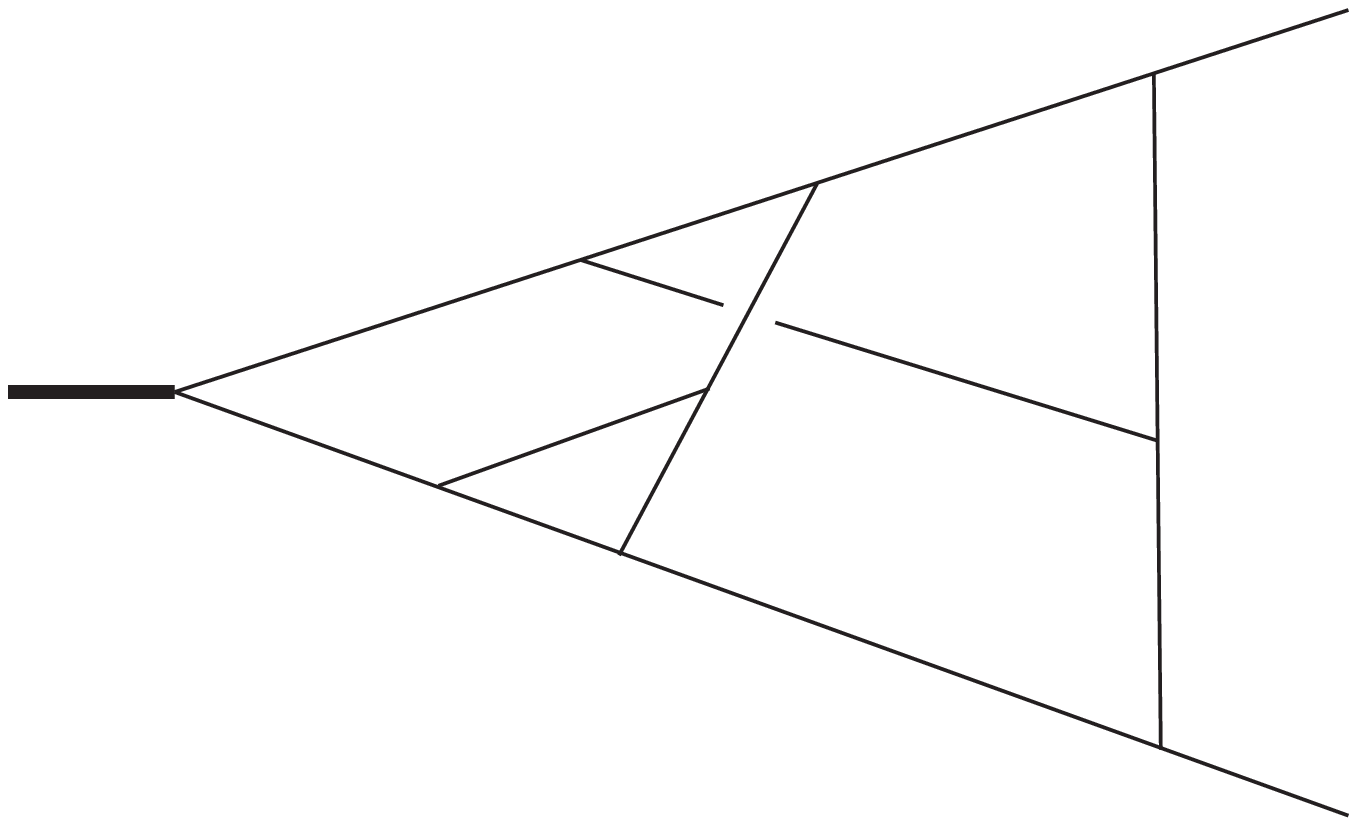} \quad \Graph{.175}{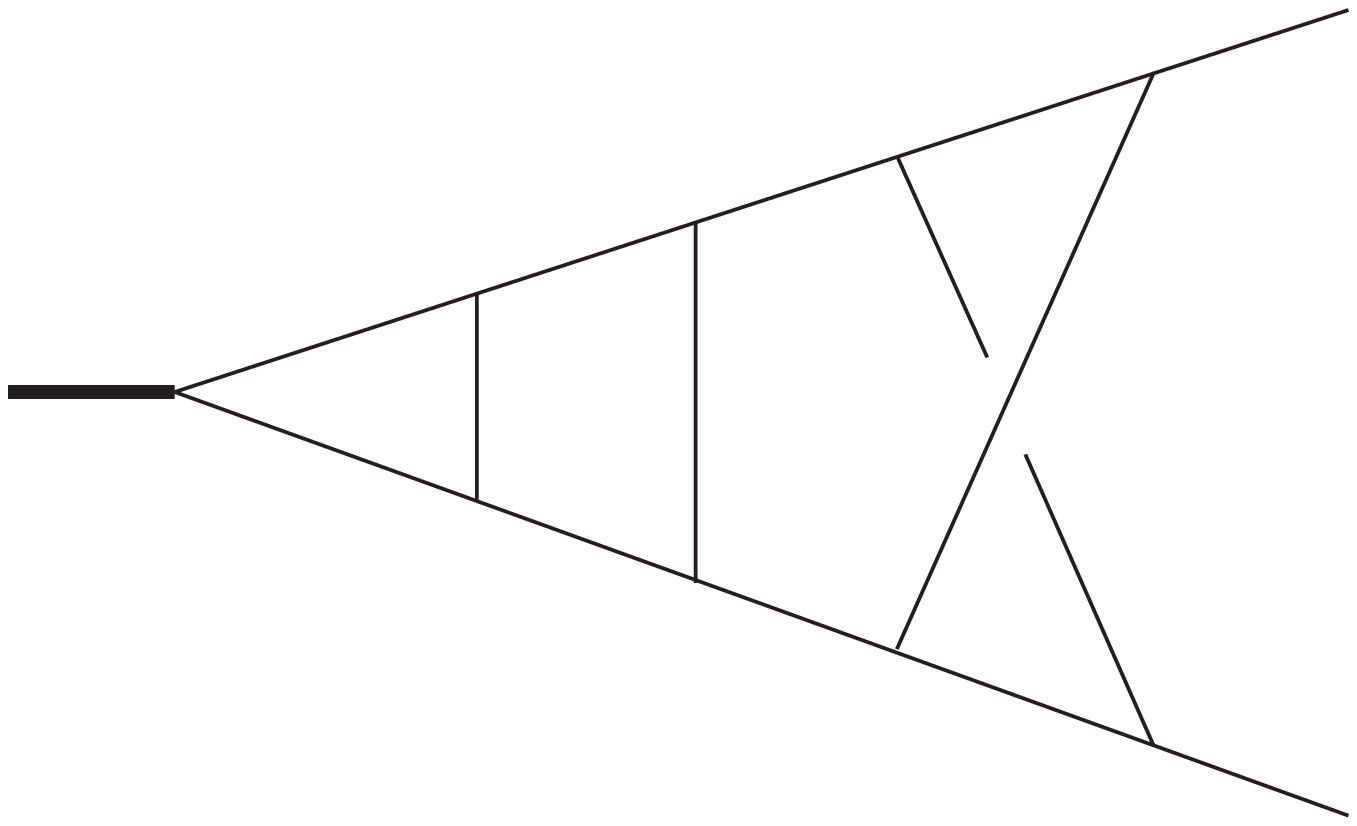}\quad \Graph{.175}{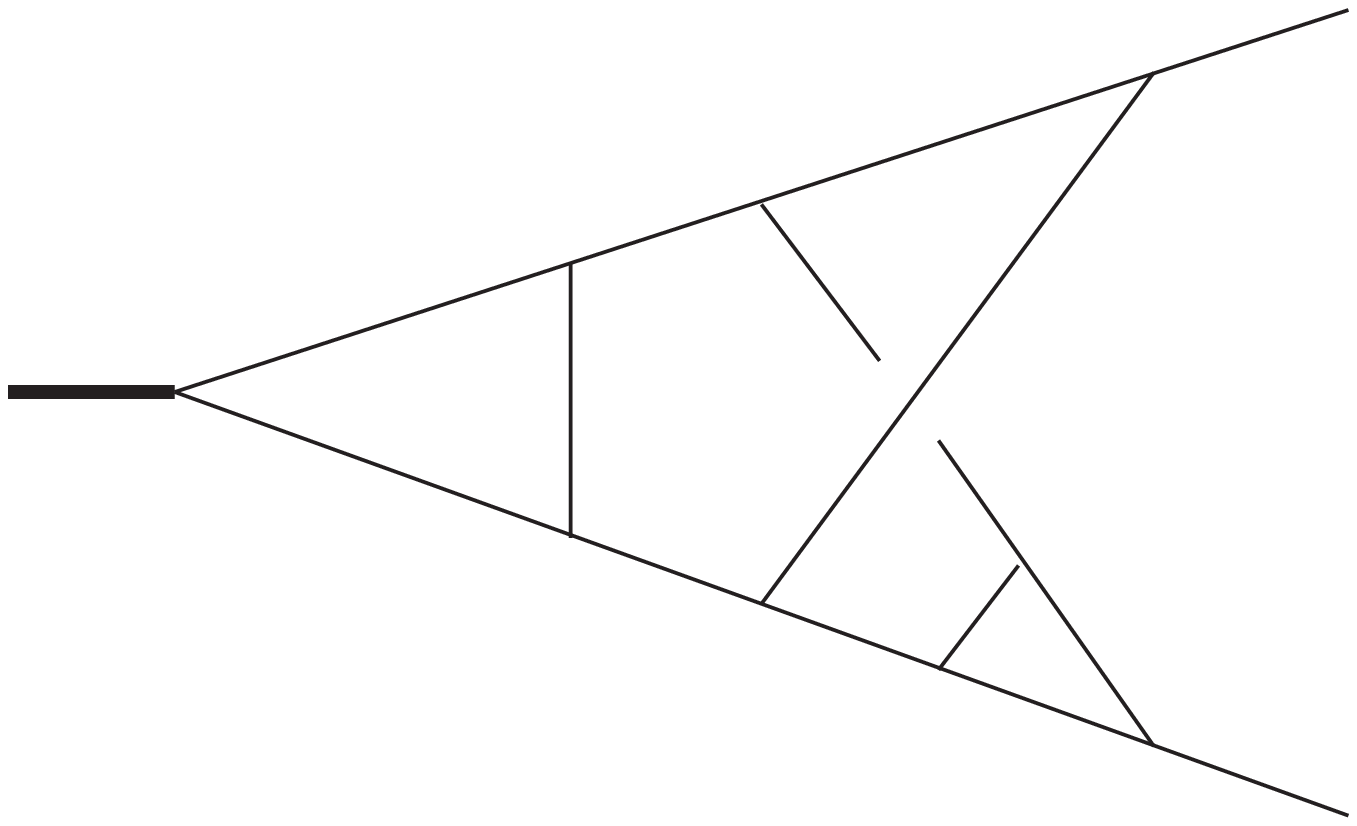} \quad \Graph{.175}{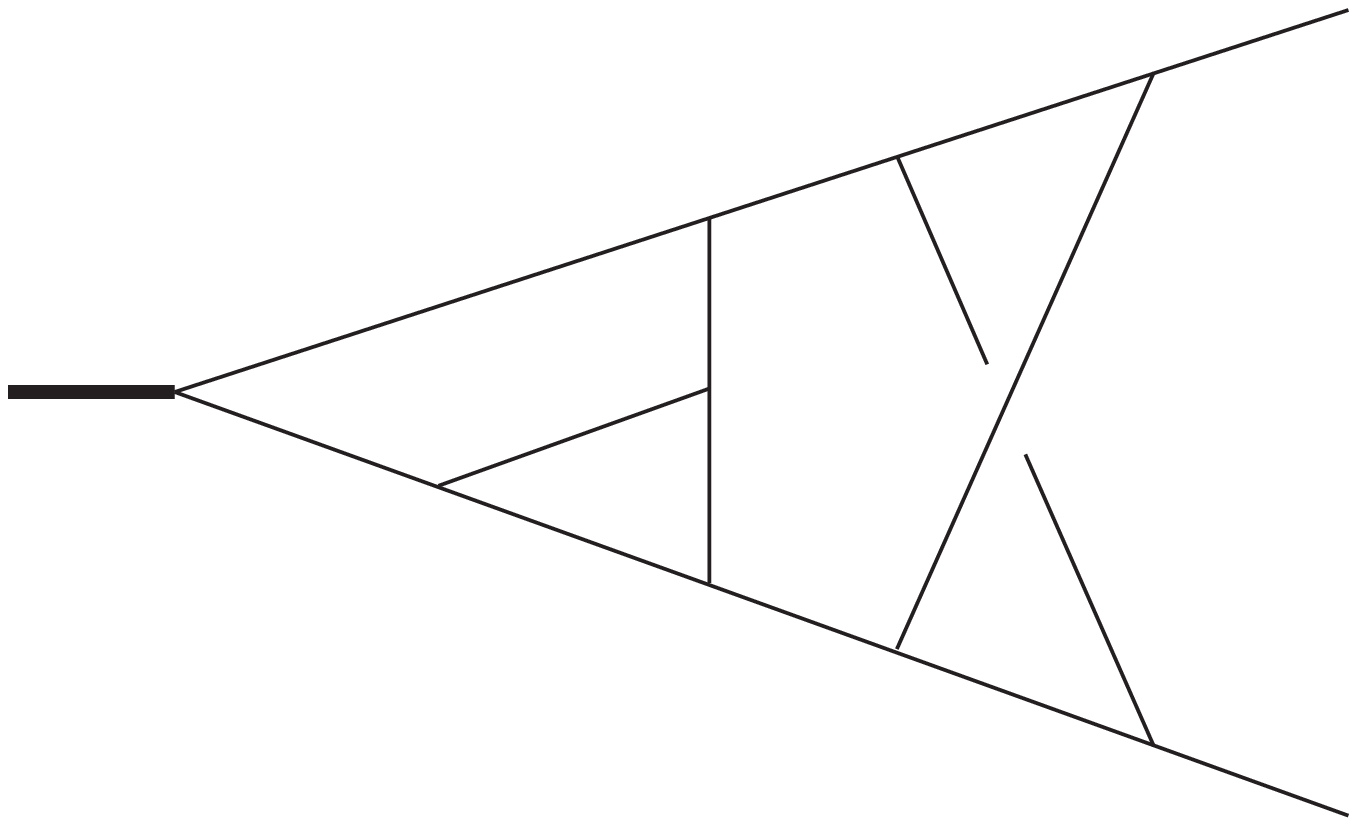}
\\
&\quad\quad \qquad \qquad \qquad \quad ~~\Graph{.175}{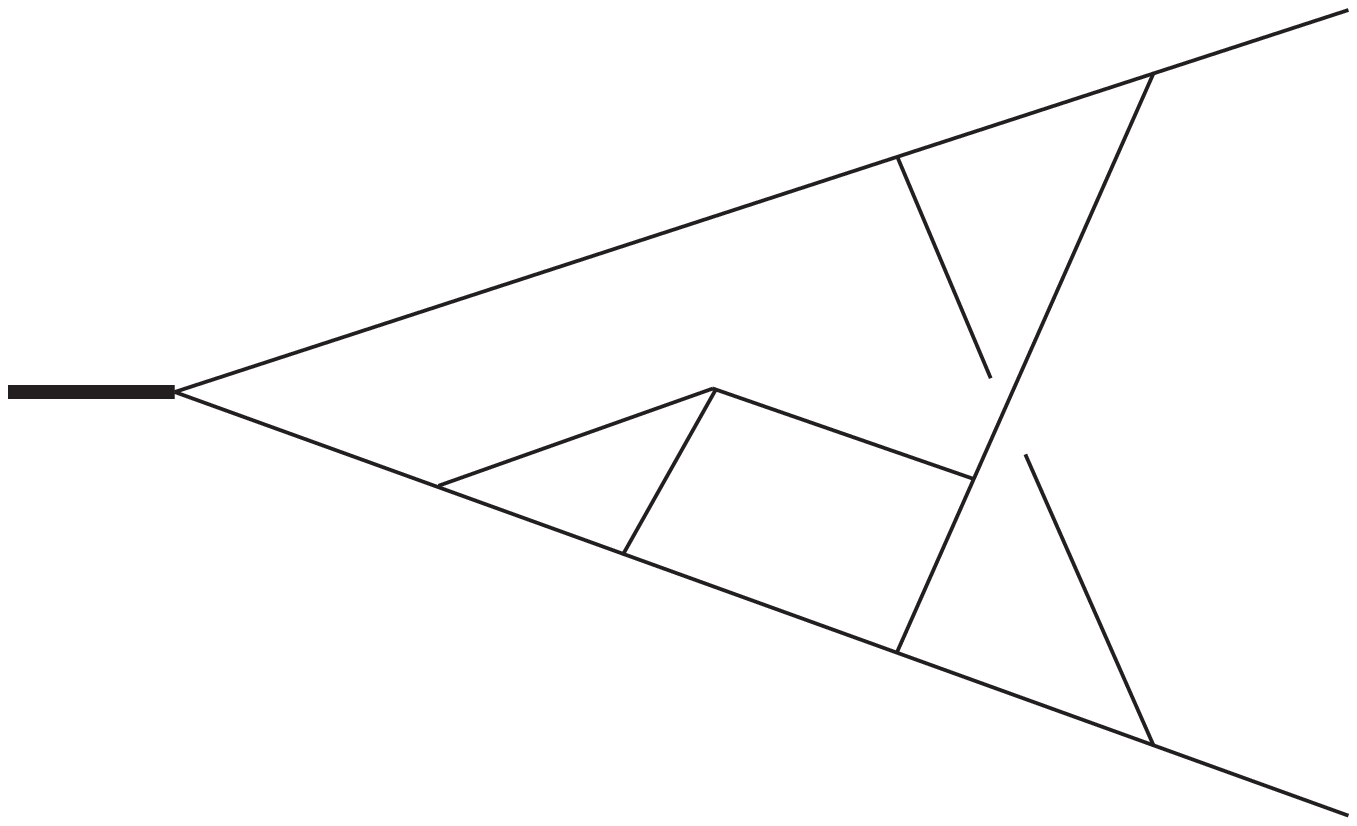} \quad \Graph{.175}{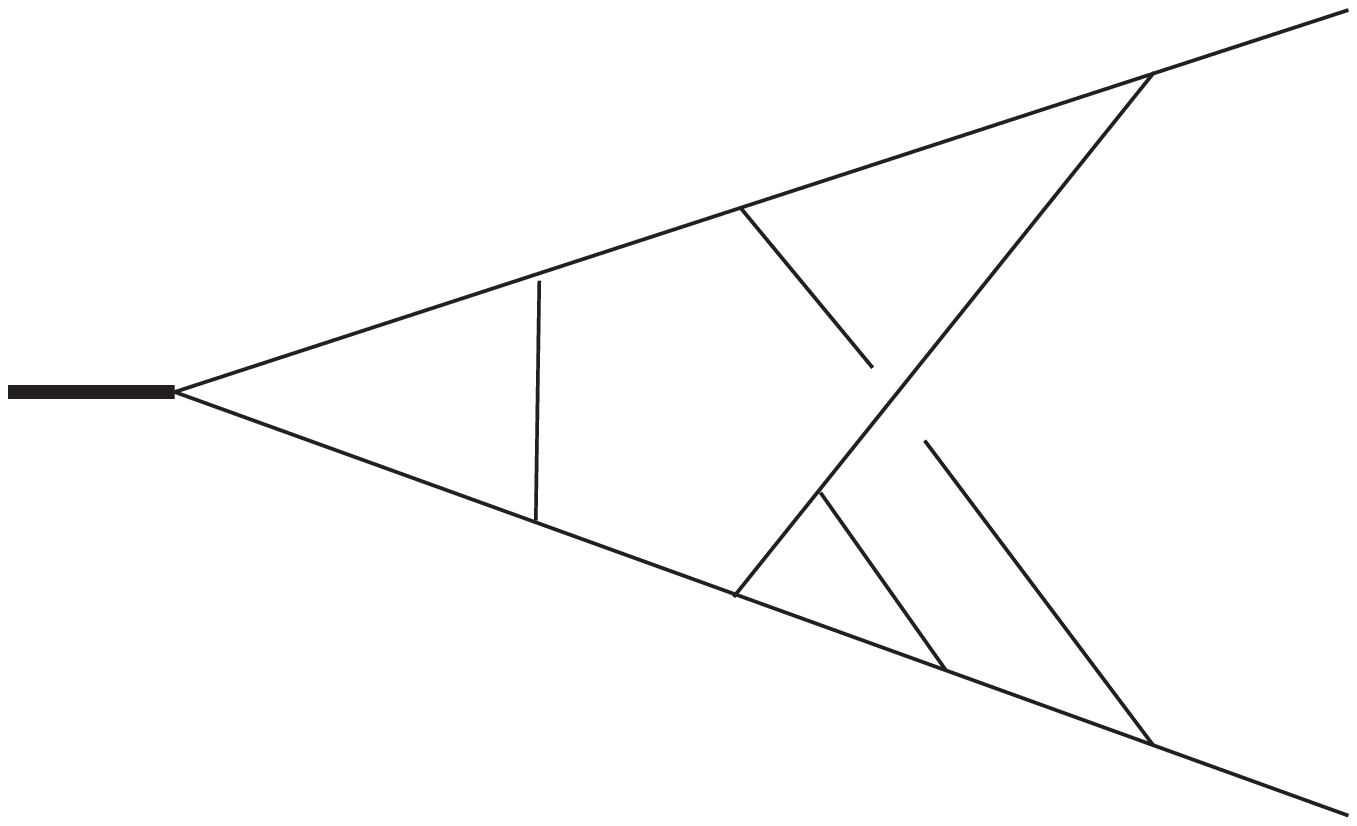} \quad \Graph{.175}{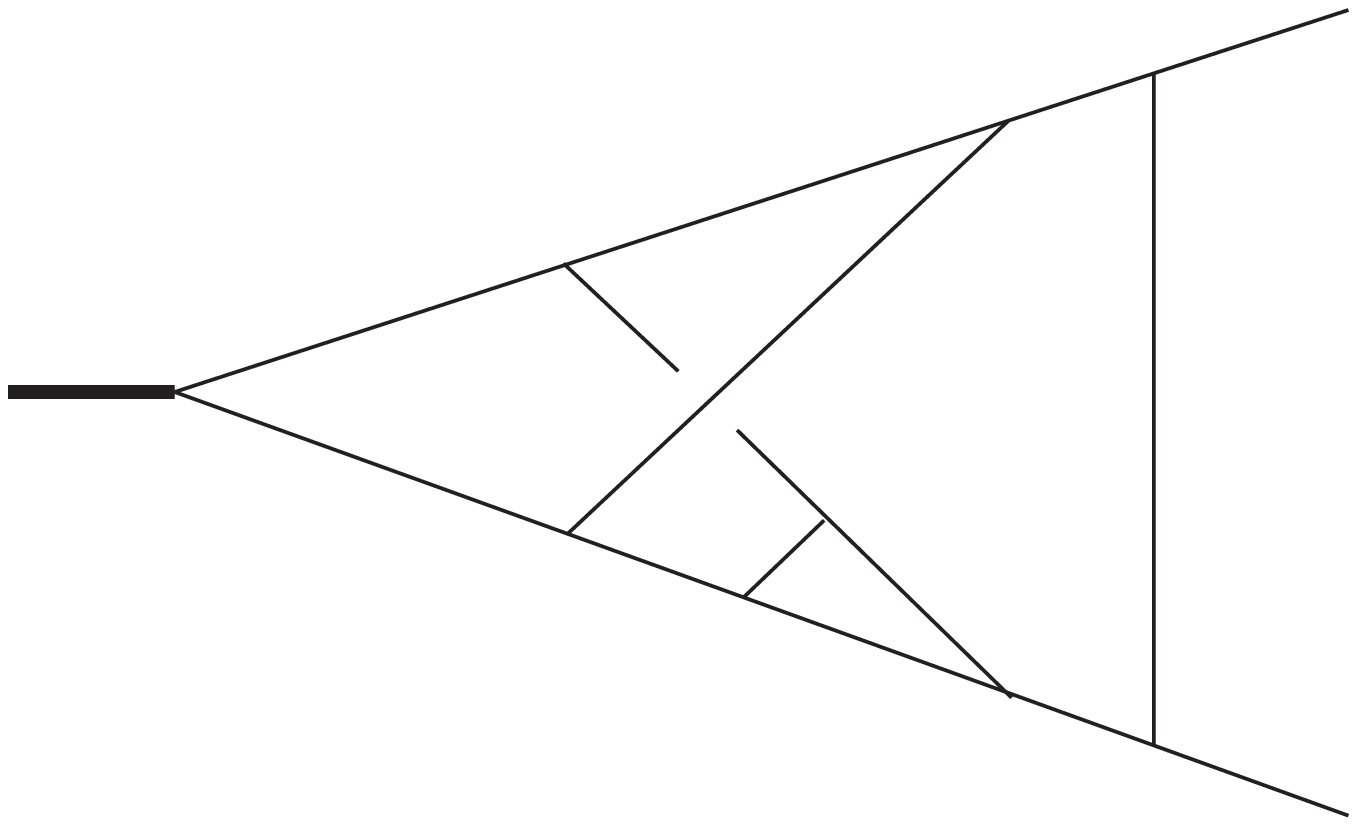} \quad \Graph{.175}{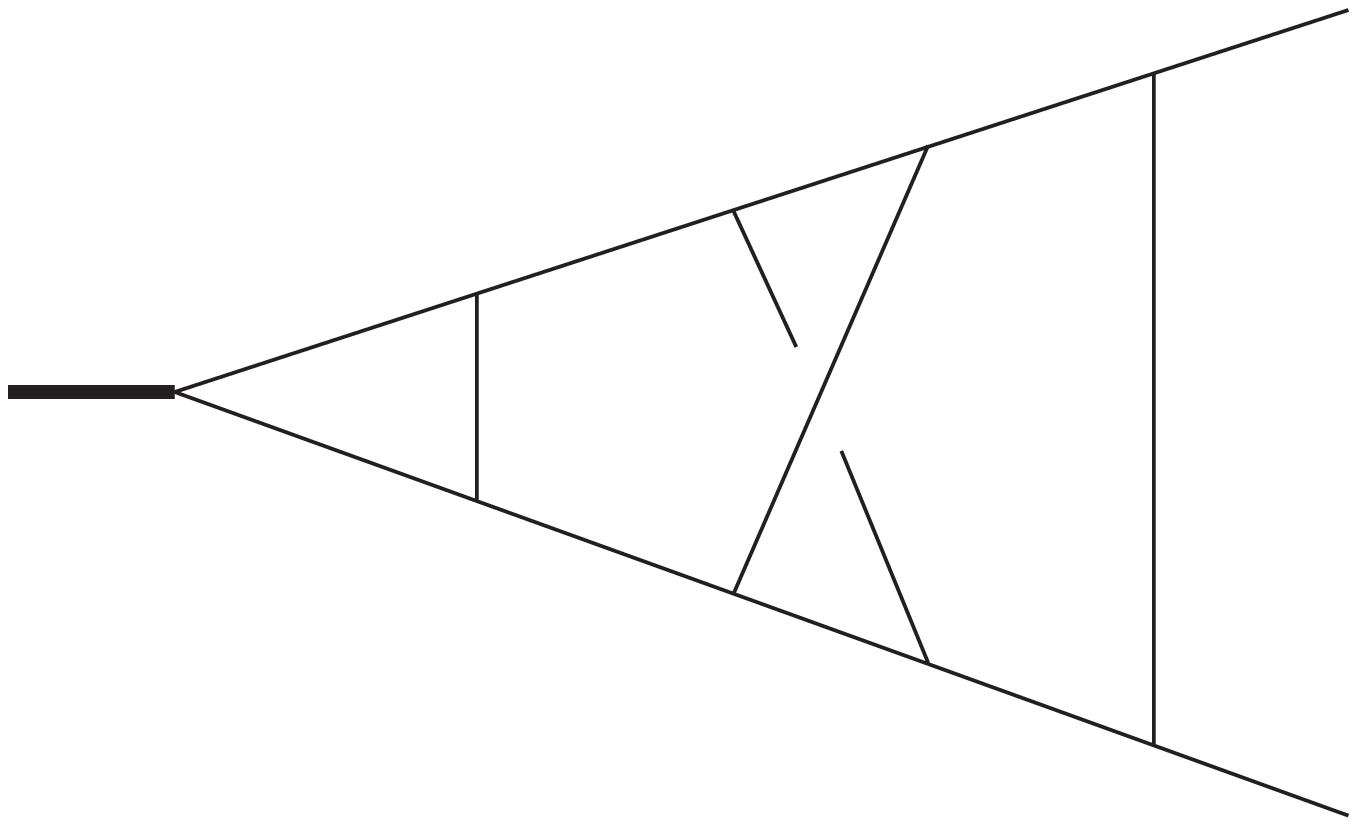}
\end{align*}
\caption{Non-planar top-level topologies contributing to our form factors. All topologies in the above except the first four turn out to be reducible. Of the four irreducible non-planar topologies, the first three are treated by us for the first time in this work and the first two are rendered linearly reducible only after a substitution of Feynman parameters. The $N_f^2$ contributions to the gluon form factor
receives contributions from all but the last three
topologies, the latter are relevant to the $N_{q \gamma} N_f$ contributions
to the quark form factor only.
The $N_f^2$ contributions to the quark form factor do not involve non-planar twelve-line topologies.}
% (see Tab. \ref{tab:diags} for further details).}
\label{fig:sampletopos}
\end{figure*}

In this paper, we continue our ongoing study of the four-loop corrections in massless Quantum Chromodynamics (QCD) to the basic quark and gluon form factors for the photon-quark-antiquark vertex and the effective Higgs-gluon-gluon vertex.
Despite a significant amount of recent attention, only partial results are available in the literature so far. The $N_f^3$ contributions to the quark and gluon form factors with three closed fermion lines were calculated, respectively, in \cite{Henn:2016men} and \cite{vonManteuffel:2016xki}. For the quark form factor, the contributions of order $N_f^2$ were computed in \cite{Lee:2017mip} and the leading color limit was obtained in
 \cite{Henn:2016men,Lee:2016ixa}. Very recently, the quartic Casimir color structure of the $N_f$ contributions to the quark form factor was calculated in \cite{Lee:2019zop} and analytic results for the corresponding contributions to the quark cusp anomalous dimension were also derived in a complementary approach in  \cite{Henn:2019rmi}.

Quartic Casimir structures in the cusp anomalous dimensions are of particular interest since they violate the Casimir scaling principle proposed in references \cite{Becher:2009cu} and \cite{Gardi:2009qi}.
The existence of such Casimir scaling violations was demonstrated by explicit numerical calculation in $\mathcal{N} = 4$ super Yang-Mills theory \cite{Boels:2017skl,Boels:2017ftb,Henn:2019rmi} and in QCD \cite{Moch:2017uml,Moch:2018wjh}. We wish to remark that, using approximation techniques \cite{vanNeerven:1999ca,vanNeerven:2000uj,vanNeerven:2000wp}, references \cite{Moch:2017uml} and \cite{Moch:2018wjh} actually delivered complete numerical results for both the quark and gluon cusp anomalous dimensions and suggested a generalization of the Casimir scaling principle.  

In this work, we substantially improve upon the current state-of-the art by providing complete analytic results for all contributions to the quark and gluon form factors with two closed fermion lines. While the $N_f^2$ contributions to the quark form factor were already known, the singlet contributions to the quark form factor of order $N_{q \gamma}N_f$,
where the photon couples to one of the two closed fermion lines, are new. We furthermore calculate the complete set of $N_f^2$ contributions to the gluon form factor for the first time,
including a determination of the $N_f^2$ terms of the gluon cusp anomalous dimension.
Due to the complexity of the relevant Feynman diagrams (see Fig. \ref{fig:sampletopos}), it was necessary to build upon specialized computational technology studied by us in earlier works \cite{vonManteuffel:2014ixa,vonManteuffel:2014qoa,vonManteuffel:2015gxa}.

One of our main calculational challenges was the reduction of four-loop, three-point Feynman integrals,
for which we employ finite field-based techniques~\cite{vonManteuffel:2014ixa,Hart2010}
implemented in the private program {\tt Finred}.
To evaluate the master integrals, we also make use of integrals with a significant number of dots (higher powers of the propagators).
For their reduction, a newly developed algorithm based on the Lee-Pomeransky parametric representation \cite{Lee:2013hzt} was used, building upon ideas discussed in \cite{Lee:2013mka,Bitoun:2017nre}.

The finite integral method of reference \cite{vonManteuffel:2015gxa} was applied to calculate all of the 163 four-loop master integrals which remain after integration by parts reduction \cite{Tkachov:1981wb,Chetyrkin:1981qh,Laporta:2001dd,vonManteuffel:2014ixa}.
Nineteen of those which we computed for the first time are of the more challenging non-planar type, free of massless one-loop bubble insertions. These include in particular master integrals for the first three top-level topologies shown in Fig. \ref{fig:sampletopos}.

To carry out the calculation, we pass to a finite integral basis and analytically integrate the $\epsilon$ expansions of the resulting integrals starting from their Feynman parametric representations. For this purpose, we employ the program {\tt HyperInt} \cite{Panzer:2014caa}.
This direct integration in terms of multiple polylogarithms requires the integrals to be {\it linearly reducible} \cite{Brown:2008um,Brown:2009ta}, a criterion which was satisfied by most of our finite integrals. Although the first two integral topologies of Fig. \ref{fig:sampletopos} were not linearly reducible initially, we were able to make simple changes of variables which rendered them so. As will be discussed below, the integration is sometimes greatly simplified by making a judicious choice of finite integral basis.

\section{Setup and Integral Reduction}

We consider the perturbative amplitudes for the decays of photons and Higgs bosons into massless partons, $\gamma^\ast(q)\rightarrow q(p_1)\bar{q}(p_2)$ and $h(q)\rightarrow g(p_1)g(p_2)$, respectively, with $p_1^2=p_2^2=0$  and $q^2=(p_1+p_2)^2$.
Interfering the bare amplitude with the tree amplitude,
summing over polarizations and color, and normalizing to the corresponding tree-level expressions, we obtain the form factors
\begin{align}
\label{eq:expbareg}
&\ff_{\rm bare}^{q,g}\left(\alpha_s^{\rm bare}, q^2, \mu_\epsilon^2, \epsilon\right) = \\
&\qquad 1 + \sum_{L = 1}^\infty \left(\frac{\alpha_s^{\rm bare}}{4\pi}\right)^L 
\left(\frac{4\pi \mu_\epsilon^2}{-q^2}\right)^{L \epsilon}
e^{-L\epsilon\gamma_E}\ff_L^{q,g}(\epsilon),\nonumber
\end{align}
where we take the $L$-loop massless QCD corrections into account.
We work in conventional dimensional regularization with the bare
strong coupling constant $\alpha_s^{\rm bare}$, the 't Hooft scale
$\mu_\epsilon$, Euler's constant $\gamma_E$, and the parameter
of dimensional regularization $\epsilon = (4-d)/2$.
The amplitudes are calculated using a general $R_\xi$ gauge for the internal gluons, with up to one power of $1-\xi$, and arbitrary reference vectors for the external gluons.
We denote the number of light quark flavors by $N_f$ and the charge-weighted sum of
the $N_f$ quark flavors normalized to the charge of the external quark $q$ by
$N_{q\gamma}\equiv  {\sum_{q^\prime} e_{q^\prime}}/e_q$.

We generate the four-loop diagrams with the program
{\tt QGraf} \cite{Nogueira:1991ex} and consider the gauge-invariant
subset with two closed fermion lines.
We match the planar and non-planar diagrams to 
 nine complete sets of eighteen denominators (integral families)
 with {\tt Reduze\;2} \cite{Bauer:2000cp,Studerus:2009ye,vonManteuffel:2012np},
where one such set of denominators may cover several twelve-line top-level topologies and equivalent
subtopologies are identified.
In total, we encounter forty-six twelve-line top-level topologies, out of which twenty-two are non-planar, see Fig.~\ref{fig:sampletopos}.
We evaluate the color algebra for a general compact simple Lie group
with {\tt Color.h} \cite{vanRitbergen:1998pn} and evaluate the Lorentz and Dirac algebra with {\tt Form\;4} \cite{Kuipers:2012rf}.

We employ the in-house program {\tt Finred} to reduce the resulting
Feynman integrals to master integrals.
For the reduction of the amplitude, we employ conventional momentum space
integration-by-parts, Lorentz, and sector symmetry identities.
For the basis change to our finite integrals, we employ first- and second-order annihilators \cite{Lee:2013mka,Bitoun:2017nre} in the Lee-Pomeransky representation.
Instead of resorting to an external computer algebra system, we
calculate the required syzygies using linear algebra methods \cite{CabarcasDing,Schabinger:2011dz} with {\tt Finred} as a linear solver.

Finite field sampling allows us to easily discard redundant equations,
which reduces the number of equations and speeds up the reduction process considerably.
Using different finite fields and different samples for $d$ allows us
to solve linear systems with 64 bit integers as coefficients
and to reconstruct the rational functions from these finite field solutions.
Robust vetos of fake reconstructions allow us to work without {\it a priori} assumptions concerning the required number of samples.
We typically need $\mathcal{O}(10^1)$ finite fields and $\mathcal{O}(10^2)$ values for $d$ for
a successful reconstruction.
We verify the reconstructed solution using five independent samples with unrelated values for $d$ and the modulus.
The reduction is run in a distributed manner and our final integral tables amount to several terabytes of compressed data.

\begin{table}[b]
\begin{tabular}{ l | c | c | c}
& $\ff_4^{q}|_{N_f^2} $ & $\ff_4^{q}|_{N_{q\gamma}N_f} $ & $\ff_4^{g}|_{N_f^2} $ \\
\hline \hline
\# diagrams & 71 & 226 & 2554 \\
\# planar twelve-line top.\ & 0 & 4 & 21 \\
\# non-planar twelve-line top.\ & 0 & 3 & 19 \\
\# non-equivalent top. in red.\ & 158 & 923 & 1781 \\
%\# integrals ($\xi \neq 1$) & 337754 & 1153514  & 20645509 \\ 
\# integrals in amp.\  ($\xi \neq 1$) & $\mathcal{O}(10^5)$ & $\mathcal{O}(10^6)$ & $\mathcal{O}(10^7)$ \\ 
%max. \# propagators & 12 & 12 & 12 \\
%max. \# diff. propagators & 11 & 12 & 12 \\
max. \# inverse propagators & 5 & 5 & 6 \\
\end{tabular}
\caption{\label{tab:diags}Complexity of various form factor contributions.}
\end{table}

Our unreduced amplitudes contain a total of 21286021 integrals
with up to twelve propagators and six inverse propagators (irreducible numerators),
for which we reconstructed a total of $\mathcal{O}(10^9)$ reduction identities in
1863 non-equivalent non-zero topologies, see Tab.~\ref{tab:diags} for more statistical data.
After insertion of the reduction identities in the amplitudes, we find 163 irreducible master integrals and observe a non-trivial cancellation
of all gauge parameter-dependent terms, something which constitutes an important sanity check on our calculation.

\section{Master Integrals}

To evaluate the master integrals using direct analytical integration, we proceed as follows.
We apply the algorithm of reference \cite{vonManteuffel:2014qoa} to generate a reasonably
long (over-complete) list of finite integrals in higher dimensions for a given sector.
For each of these integrals, we evaluate the leading term by direct integration of the
Feynman parametric representation using {\tt HyperInt}.
To avoid the evaluation of many terms in the $\epsilon$ expansion,
it is often desirable to select a basis of finite integrals with high maximal transcendental weights at leading order in $\epsilon$.
Using dimensional recurrence \cite{Tarasov:1996br,Lee:2012cn} and dedicated reduction identities, we express the conventional master integrals in $4 - 2\epsilon$ dimensions in terms of the new finite master integrals and their subsectors, which should be thought of as being known in a bottom-up approach. 
In this way, we obtain the reduced form factors directly in terms of finite master integrals and subsequently insert the analytical solution for their Taylor expansion at $\epsilon=0$.

One advantage of this choice of basis is that the finite integrals
enter the $\epsilon$ expansion of amplitudes at higher orders than their conventional
counterparts.
To illustrate this effect, let us express explicit integration results
using different choices for the master integrals.
For example, for the first integral topology of Fig. \ref{fig:sampletopos}, we find in the conventional basis
\begin{align}
\label{eq:J_12_4095}
&\figgraph{.185}{J_12_4095}{4} = \frac{1}{\epsilon^8}\left(\frac{1}{72}\right) +\frac{1}{\epsilon^6}\left(-\frac{11}{72}\zeta_2+\frac{13}{72}\right)
\nonumber\\&
+\frac{1}{\epsilon^5}\left(-\frac{191}{72}\zeta_3-\zeta_2-\frac{17}{24}\right)+\frac{1}{\epsilon^4}\left(-\frac{2779}{360}\zeta_2^2-15\zeta_3
\right.\nonumber\\&\left.
+\frac{8}{3}\zeta_2+\frac{5}{24}\right)+\frac{1}{\epsilon^3}\left(-\frac{469}{8}\zeta_5-\frac{205}{36}\zeta_2\zeta_3-27\zeta_2^2+\frac{251}{6}\zeta_3
\right.\nonumber\\&\left.
-\frac{65}{3}\zeta_2+\frac{451}{24}\right)
+\frac{1}{\epsilon^2}\left(\frac{21269}{72}\zeta_3^2-\frac{140807}{1260}\zeta_2^3+545\zeta_5
\right.\nonumber\\&\left.
-222 \zeta_2\zeta_3+\frac{1067}{10}\zeta_2^2-\frac{1399}{6}\zeta_3+\frac{580}{3}\zeta_2-\frac{4459}{24}\right)
\nonumber\\&
+\frac{1}{\epsilon}\left(- \frac{11417}{24}\zeta_7 - \frac{9445}{12}\zeta_2\zeta_5 
+ \frac{181147}{180}\zeta_2^2 \zeta_3  + 2871 \zeta_3^2
\right.\nonumber\\&\left.
+\frac{34414}{105}\zeta_2^3 + \frac{561}{2}\zeta_5 + \frac{1720}{3} \zeta_2 \zeta_3- \frac{17993}{30} \zeta_2^2 + \frac{10505}{6} \zeta_3
\right.\nonumber\\&\left.
 - \frac{4361}{3}\zeta_2+\frac{31003}{24}\right) - \frac{86152}{15} \zeta_{5, 3} + \frac{47869}{12} \zeta_3 \zeta_5 
\nonumber\\&
+ \frac{31999}{36} \zeta_2 \zeta_3^2 + \frac{33270103}{21000} \zeta_2^4+\frac{134135}{2} \zeta_7 - 
 8110 \zeta_2 \zeta_5 
 \nonumber\\&
 + \frac{23202}{5} \zeta_2^2 \zeta_3 + \frac{948274}{315} \zeta_2^3 + \frac{8717}{3} \zeta_3^2  - \frac{43085}{2} \zeta_5 
 \nonumber\\&
 + \frac{5230}{3} \zeta_2 \zeta_3  + \frac{117067}{30} \zeta_2^2 - \frac{76915}{6} \zeta_3+ \frac{28828}{3} \zeta_2 
 \nonumber\\&
 -\frac{183475}{24}+\mathcal{O}\left(\epsilon\right),
\end{align}
which contributes to all terms in the $\epsilon$ expansions of the bare form factors through to $\mathcal{O}\left(\epsilon^0\right)$.
Note that, in Eq. (\ref{eq:J_12_4095}) and throughout this work, we adopt the conventions and definitions of reference \cite{vonManteuffel:2015gxa} for our Feynman integrals and multiple zeta values.

Now, if one replaces the above master integral with a suitable finite representative, one can make explicit that \emph{no} evaluation of the new, finite master is required to extract the finite parts of the form factor contributions discussed in this work. This is simply because the finite parts of the contributions of order $N_f^2$ and $N_{q\gamma}N_f$ turn out to have a maximal weight of six (see Eqs. (\ref{eq:ffquark}), (\ref{eq:ffquarksinglet}), and (\ref{eq:ffgluon}) below), but the leading term in the $\epsilon$ expansion of the finite master integral for this sector may be consistently chosen to have a maximal weight of seven. This feature is particularly compelling for the example considered, because its finite integrals were only accessible to {\tt HyperInt} after we made a change of variable of the form $x_i = x_j x_k x_i^\prime/x_\ell$ for one of the Feynman parameters, $x_i$.

Let us elaborate further on what we require from our finite basis integrals. In most cases, it is very useful to pick finite integrals which faithfully map the weights of the underlying multiple zeta values when passing from the preferred finite basis back to the conventional basis. That is to say, we try whenever possible to ensure that a maximal weight of $w$ at leading order in $\epsilon$ on the finite integral side implies complete weight $w$ information on the conventional integral side. This is not automatic and is closely related to the problem of finding an integral basis for a given sector which is simultaneously finite and uniform weight \cite{Schabinger:2018dyi}. As a dramatic example, consider the conventional master integrals
\begin{align}
\label{eq:B_10_20339}
&\figgraph{.15}{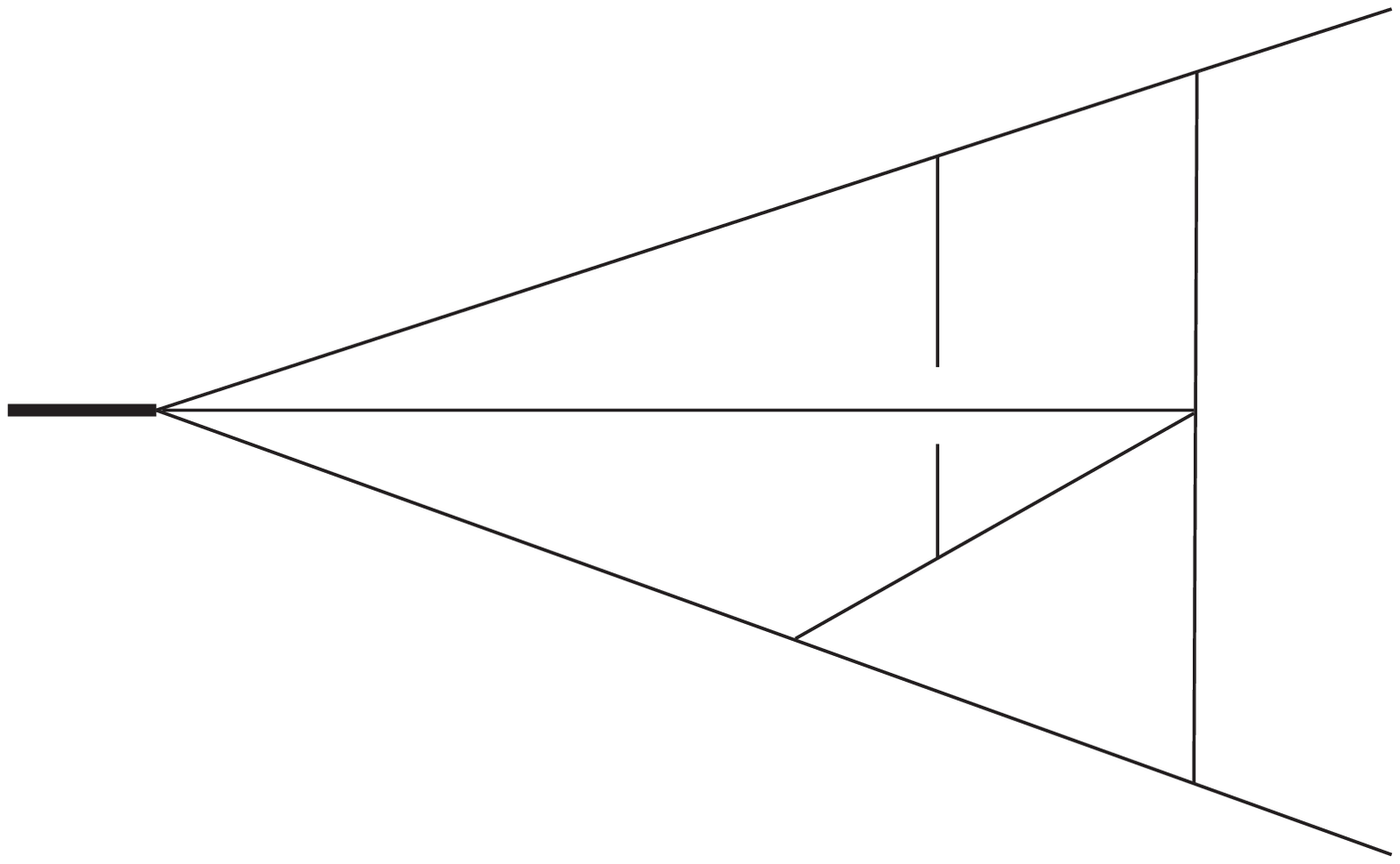}{4} =  \frac{1}{\epsilon^8}\left(\frac{1}{144}\right) +\frac{1}{\epsilon^6}\left(- \frac{1}{24}\zeta_2\right)
\nonumber\\&
+\frac{1}{\epsilon^5}\left(- \frac{29}{24}\zeta_3\right) +\frac{1}{\epsilon^4}\left(- \frac{71}{16}\zeta_2^2\right) + \frac{1}{\epsilon^3}\left(
  - \frac{1819}{24}\zeta_5
\right.\nonumber\\&\left.
+\frac{23}{6} \zeta_2 \zeta_3\right) 
  + \frac{1}{\epsilon^2}\left(\frac{1285}{24} \zeta_3^2-\frac{80579}{1008} \zeta_2^3\right) 
\nonumber\\&
  + \frac{1}{\epsilon}\left( - \frac{434203}{192} \zeta_7 + \frac{7139}{24} \zeta_2 \zeta_5+\frac{54139}{120} \zeta_2^2 \zeta_3\right)  + \frac{2023}{12} \zeta_{5,3}
\nonumber\\&
 + \frac{30581}{4} \zeta_3 \zeta_5+ 
 \frac{6829}{24} \zeta_2 \zeta_3^2- \frac{45893321}{100800} \zeta_2^4 
 +\mathcal{O}\left(\epsilon\right)
\end{align}
and
\begin{align}
\label{eq:B_10_20339_dot}
&\figgraph{.15}{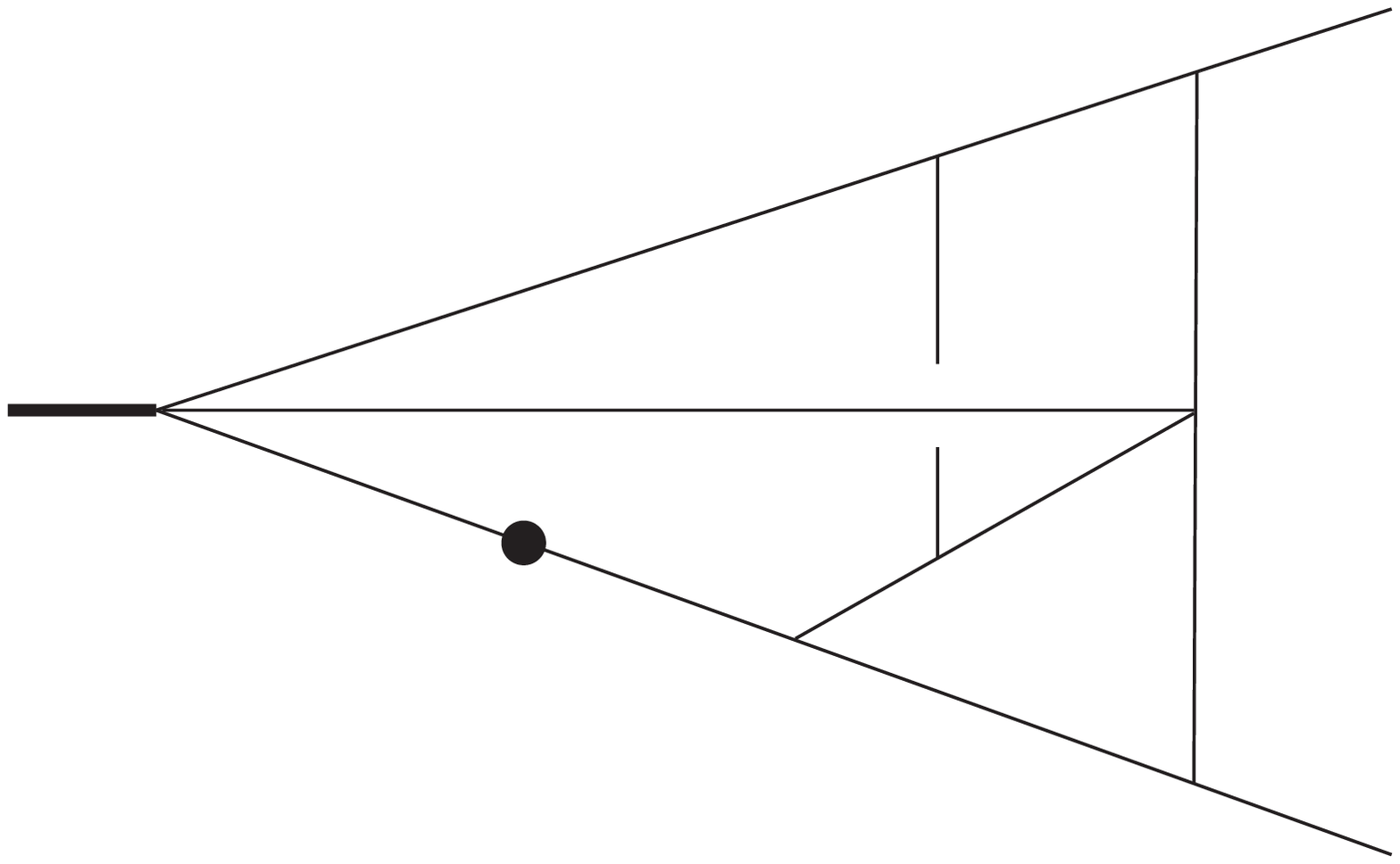}{4} = \frac{1}{\epsilon^8}\left(-\frac{1}{144}\right)+\frac{1}{\epsilon^7}\left(-\frac{1}{12}\right)
\nonumber\\&
+\frac{1}{\epsilon^6}\left(\frac{1}{24}\zeta_2-\frac{7}{36}\right) +\frac{1}{\epsilon^5} \left(\frac{29}{24} \zeta_3 + \frac{1}{2}\zeta_2 -\frac{1}{72}\right) 
\nonumber\\&
+ \frac{1}{\epsilon^4}\left(\frac{71}{16} \zeta_2^2 + \frac{29}{2} \zeta_3+ \frac{39}{16}\zeta_2+\frac{335}{144}\right) + \frac{1}{\epsilon^3}\left(
 \frac{1819}{24} \zeta_5 
\right.\nonumber\\&\left.
 - \frac{23}{6} \zeta_2 \zeta_3 + \frac{213}{4} \zeta_2^2 + \frac{1211}{24} \zeta_3  + \frac{431}{48}\zeta_2 + \frac{47}{18}\right) 
 \nonumber\\&
 + \frac{1}{\epsilon^2} \left(- \frac{1285}{24}\zeta_3^2 + \frac{80579}{1008}\zeta_2^3 + \frac{1819}{2} \zeta_5 - 46 \zeta_2 \zeta_3 
 \right.\nonumber\\&\left.
 + \frac{25787}{160} \zeta_2^2 + \frac{417}{8} \zeta_3 - \frac{1175}{48} \zeta_2 -\frac{7277}{72}\right) + 
 \frac{1}{\epsilon} \left(\frac{434203}{192} \zeta_7 
 \right.\nonumber\\&\left.
 - \frac{7139}{24} \zeta_2 \zeta_5 - \frac{54139}{120} \zeta_2^2 \zeta_3 - \frac{1285}{2} \zeta_3^2 + \frac{80579}{84}\zeta_2^3 
 \right.\nonumber\\&\left.
 + \frac{5571}{2} \zeta_5 - \frac{9005}{24} \zeta_2 \zeta_3 
   + \frac{967}{480} \zeta_2^2  - \frac{4045}{8} \zeta_3- \frac{733}{24} \zeta_2
 \right.\nonumber\\&\left.  
+ \frac{57635}{72}\right) - \frac{2023}{12} \zeta_{5,3}-\frac{30581}{4} \zeta_3 \zeta_5 - \frac{6829}{24} \zeta_2 \zeta_3^2 
\nonumber\\&
+ \frac{45893321}{100800} \zeta_2^4 + \frac{434203}{16} \zeta_7 - 
 \frac{7139}{2} \zeta_2 \zeta_5 - \frac{54139}{10} \zeta_2^2 \zeta_3  
 \nonumber\\&
 - \frac{10706}{3} \zeta_3^2 + \frac{7987951}{3360} \zeta_2^3  + \frac{1309}{12} \zeta_5 -  \frac{30317}{24} \zeta_2 \zeta_3 
 \nonumber\\&
 - \frac{43847}{96} \zeta_2^2 + \frac{32335}{24} \zeta_3  + \frac{2553}{4} \zeta_2 - \frac{334727}{72} + \mathcal{O}\left(\epsilon\right). \raisetag{16pt}
\end{align}

With the master integrals above, the reduced integrand for the $N_f^2$ four-loop gluon form factor would actually require the unknown $\mathcal{O}\left(\epsilon\right)$ terms of Eqs. (\ref{eq:B_10_20339}) and (\ref{eq:B_10_20339_dot}) to extract final results through to $\mathcal{O}\left(\epsilon^0\right)$. On the other hand, a suitable finite integral basis including
\begin{align}
\label{eq:B_10_20339_fin_1}
&\figgraph{.15}{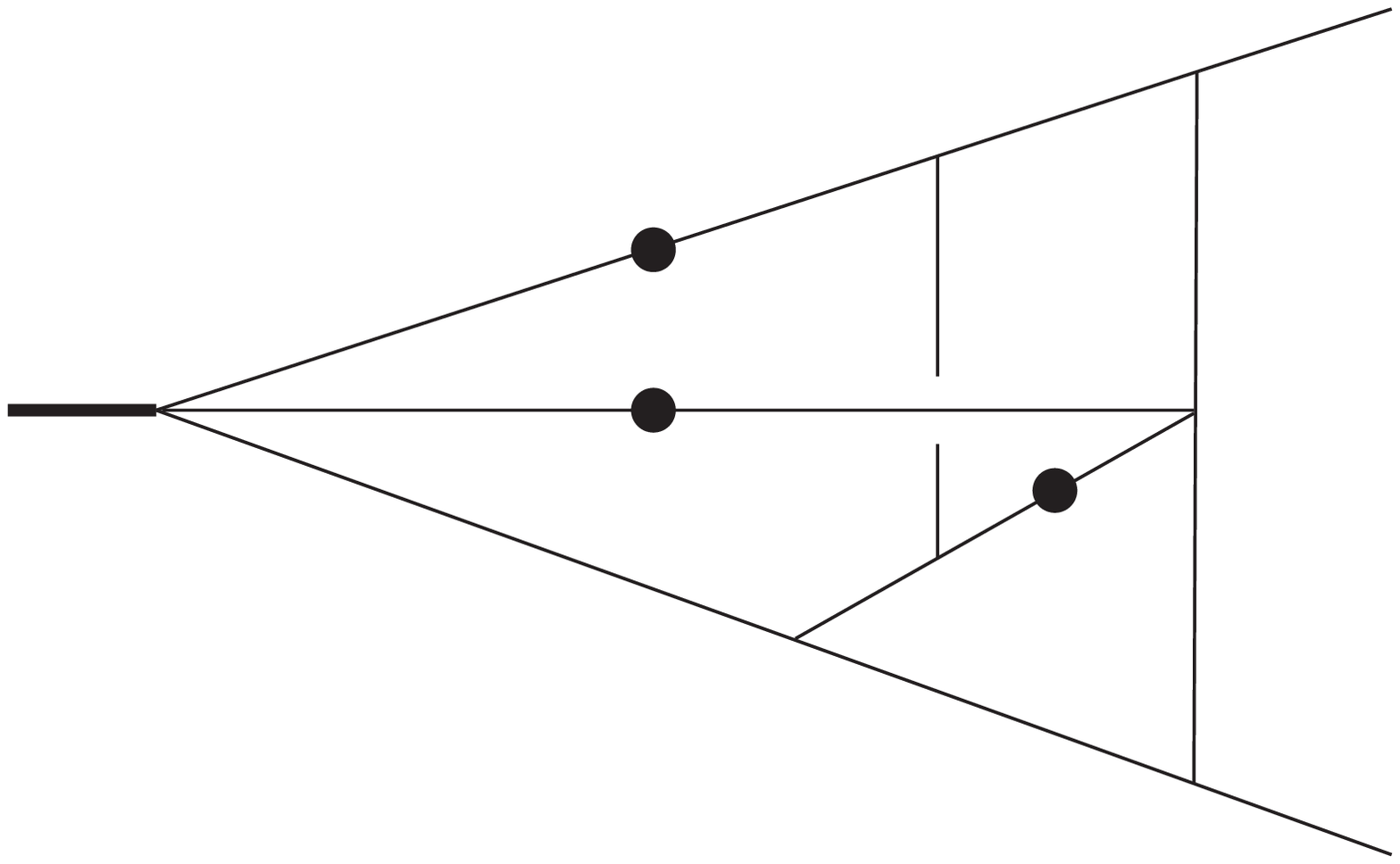}{6} = - \frac{3}{2} \zeta_3^2 - \frac{4}{3} \zeta_2^3+10 \zeta_5 + 2 \zeta_2 \zeta_3  - \frac{1}{5}\zeta_2^2
\nonumber \\&
- 6 \zeta_3 +\epsilon \left(- \frac{273}{8}\zeta_7- \frac{155}{4} \zeta_2 \zeta_5 + \frac{58}{5} \zeta_2^2 \zeta_3+ \frac{107}{4} \zeta_3^2
\right.\nonumber\\&\left.
+ \frac{419}{60} \zeta_2^3+ \frac{311}{4} \zeta_5+ \frac{55}{2} \zeta_2 \zeta_3-\frac{71}{10} \zeta_2^2  - 117 \zeta_3\right) 
\nonumber \\&
+  \epsilon^2 \left(\frac{107}{4} \zeta_{5,3} + \frac{99}{4} \zeta_3 \zeta_5- 
    \frac{135}{4} \zeta_2 \zeta_3^2 - \frac{50587}{1050} \zeta_2^4
\right.\nonumber\\&\left.    
+ \frac{4245}{8}\zeta_7- \frac{545}{2} \zeta_2 \zeta_5 + \frac{1201}{10} \zeta_2^2 \zeta_3+ \frac{1303}{4} \zeta_3^2+ \frac{13491}{140} \zeta_2^3
\right.\nonumber\\&\left.
+ \frac{1837}{4} \zeta_5+ \frac{581}{2} \zeta_2 \zeta_3     -\frac{569}{5} \zeta_2^2 - 1542 \zeta_3 \right)+\mathcal{O}\left(\epsilon^3\right)
\end{align}
offers the possibility to completely avoid evaluating anything but the weight six, leading-order term of Eq. (\ref{eq:B_10_20339_fin_1}). For the majority of the 151 irreducible topologies relevant to the $N_f^2$ gluon form factor, we were able to find finite basis integrals which allowed us to avoid computing spurious orders of their $\epsilon$ expansions.

\section{Results and Cross-Checks}

With the techniques described above, we find for the non-singlet quark form factor
\begin{align}
\label{eq:ffquark}
&\ff_4^{q}(\epsilon)\Big|_{N_f^2} = \txblu{C_F^2} \bigg[
  \frac{1}{\epsilon^6}\left(\frac{41}{162}\right)
  +\frac{1}{\epsilon^5}\left(\frac{574}{243}\right)
  +\frac{1}{\epsilon^4}\left(\frac{73}{81}\zeta_2
  \right. \nonumber\\& \left. 
+\frac{835}{54}\right)+\frac{1}{\epsilon^3}\left(-\frac{2620}{243}\zeta_3 + \frac{3016}{243}\zeta_2+\frac{176390}{2187}\right)
  \nonumber \\&
+\frac{1}{\epsilon^2}\left( -\frac{1072}{405} \zeta_2^2 -\frac{70444}{729}\zeta_3 + \frac{2693}{27}\zeta_2+ \frac{2404400}{6561}\right)
  \nonumber\\&
+\frac{1}{\epsilon}\left(\frac{8948}{405}\zeta_5 + \frac{7808}{243}\zeta_2\zeta_3+\frac{2356}{243}\zeta_2^2-\frac{69377}{81}\zeta_3
\right.\nonumber\\&\left.
+ \frac{1354325}{2187}\zeta_2+\frac{78894607}{52488}\right)+\frac{689582}{729}\zeta_3^2 + \frac{191252}{945}\zeta_2^3
\nonumber\\&
+\frac{187364}{1215}\zeta_5+\frac{177800}{729}\zeta_2\zeta_3 + \frac{4777}{135}\zeta_2^2-\frac{90719803}{13122}\zeta_3
\nonumber\\&
+ \frac{44208841}{13122}\zeta_2+\frac{5325319081}{944784}+\mathcal{O}(\epsilon) \bigg]
\nonumber\\&
+\txblu{C_A C_F} \bigg[
  \frac{1}{\epsilon^5}\left(-\frac{11}{18}\right)
  +\frac{1}{\epsilon^4}\left(\frac{1}{3}\zeta_2 -\frac{395}{54}\right)
  +\frac{1}{\epsilon^3}\left(\frac{19}{3}\zeta_3 
  \right. \nonumber\\& \left.
- 5\zeta_2-\frac{75619}{1296}\right)+\frac{1}{\epsilon^2}\left( \frac{39}{5} \zeta_2^2  +\frac{2170}{27}\zeta_3 - \frac{2044}{27}\zeta_2 
\right. \nonumber\\ & \left.
- \frac{2953141}{7776}\right)+\frac{1}{\epsilon}\left(\frac{784}{9}\zeta_5 - \frac{412}{9}\zeta_2\zeta_3+\frac{1006}{45}\zeta_2^2
\right.\nonumber\\&\left. 
+\frac{61459}{81}\zeta_3 - \frac{424399}{648}\zeta_2-\frac{102630137}{46656}\right)-\frac{1714}{3}\zeta_3^2 
\nonumber\\&
- \frac{2836}{315}\zeta_2^3 +\frac{150886}{135}\zeta_5-\frac{436}{3}\zeta_2\zeta_3 - \frac{3722}{135}\zeta_2^2
\nonumber\\&
+\frac{2897315}{486}\zeta_3 - \frac{5825827}{1296}\zeta_2-\frac{3325501813}{279936}
  +\mathcal{O}(\epsilon) \bigg],
\end{align}
for the singlet quark form factor
\begin{align}
\label{eq:ffquarksinglet}
&\ff_4^{q}(\epsilon)\Big|_{N_{q\gamma}N_f} = \txblu{\frac{d_F^{abc}d_F^{abc}}{N_F}} \bigg[
  \frac{1}{\epsilon}\left(\frac{1280}{3}\zeta_5 +\frac{32}{5}\zeta_2^2 -\frac{224}{3}\zeta_3
  \right.\nonumber\\&\left.  
- 160\zeta_2-64\vphantom{\frac{1280}{3}}\right)+2304\zeta_3^2 + \frac{545792}{945}\zeta_2^3 -\frac{8512}{9}\zeta_5
\nonumber\\&
-\frac{1888}{3}\zeta_2\zeta_3 + \frac{12448}{45}\zeta_2^2
-\frac{3520}{9}\zeta_3 - \frac{16928}{9}\zeta_2-\frac{11296}{9}
\nonumber \\&
+\mathcal{O}(\epsilon) \bigg],
\end{align}
and for the gluon form factor
\begin{align}
\label{eq:ffgluon}
&\ff_4^g(\epsilon)\Big|_{N_f^2} = \txblu{C_A^2}\bigg[
  \frac{1}{\epsilon^6}\left(\frac{41}{162}\right)
  +\frac{1}{\epsilon^5}\left(\frac{113}{486}\right)
  +\frac{1}{\epsilon^4}\left(-\frac{128}{81}\zeta_2
  \right. \nonumber\\&
  \left. -\frac{151}{27}\right)
  +\frac{1}{\epsilon^3}\left(-\frac{5167}{243}\zeta_3 + \frac{799}{243}\zeta_2-\frac{1085641}{34992}\right)
  \nonumber \\&
  +\frac{1}{\epsilon^2}\left( -\frac{7087}{405} \zeta_2^2 
 -\frac{35848}{729}\zeta_3 + \frac{425}{9}\zeta_2- \frac{1144163}{209952}\right)
  \nonumber\\&
  +\frac{1}{\epsilon}\left(-\frac{73282}{405}\zeta_5 + \frac{26396}{243}\zeta_2\zeta_3-\frac{37894}{1215}\zeta_2^2+\frac{28705}{324}\zeta_3
  \right.\nonumber\\&\left.
  + \frac{2504629}{17496}\zeta_2+\frac{541388327}{419904}\right)+\frac{717266}{729}\zeta_3^2 + \frac{58858}{945}\zeta_2^3\nonumber\\&
  -\frac{62671}{1215}\zeta_5-\frac{87329}{1458}\zeta_2\zeta_3 + \frac{380701}{3240}\zeta_2^2+\frac{27036815}{52488}\zeta_3
\nonumber\\&
- \frac{53253361}{104976}\zeta_2+\frac{110016540845}{7558272}
  +\mathcal{O}(\epsilon)\bigg] 
  \nonumber\\&
  +\txblu{C_A C_F}\bigg[
\frac{1}{\epsilon^4}\left(-\frac{7}{3}\right)
  +\frac{1}{\epsilon^3}\left(26\zeta_3 -\frac{1115}{36}\right) +\frac{1}{\epsilon^2}\left( \frac{806}{45} \zeta_2^2
\right.  \nonumber \\&\left.
+\frac{580}{9}\zeta_3 + \frac{148}{9}\zeta_2- \frac{12749}{108}\right)
+\frac{1}{\epsilon}\left(\frac{442}{9}\zeta_5 - \frac{1132}{9}\zeta_2\zeta_3
\right.  \nonumber\\&\left.
+\frac{1784}{45}\zeta_2^2-\frac{5797}{27}\zeta_3+ \frac{1475}{18}\zeta_2+\frac{642623}{1296}\right)-\frac{4354}{3}\zeta_3^2
\nonumber \\&
- \frac{143524}{945}\zeta_2^3
-\frac{47948}{27}\zeta_5-\frac{3688}{9}\zeta_2\zeta_3 - \frac{13820}{81}\zeta_2^2
\nonumber\\&
-\frac{538853}{162}\zeta_3- \frac{27629}{81}\zeta_2+\frac{3697777}{324}
  +\mathcal{O}(\epsilon) \bigg]
  \nonumber\\&
  +\txblu{C_F^2}\bigg[
\frac{1}{\epsilon^2}\left(\frac{1}{2}\right)+\frac{1}{\epsilon} \left(320\zeta_5 -236\zeta_3 -\frac{74}{3}\right)+\frac{2432}{3}\zeta_3^2
\nonumber \\&
+ \frac{32512}{135} \zeta_2^3 +3360\zeta_5 +\frac{128}{3} \zeta_3 \zeta_2-\frac{484}{3}\zeta_2^2 -\frac{36260}{9}\zeta_3 
\nonumber \\&  +\frac{439}{9}\zeta_2 -\frac{80281}{216} + \mathcal{O}\left(\epsilon\right) \bigg]
\nonumber \\&
+\txblu{\frac{d_F^{abcd}d_F^{abcd}}{N_A}}\bigg[
\frac{1}{\epsilon} \left(128\zeta_3 -\frac{176}{3}\right) +512\zeta_3^2-960\zeta_5 
\nonumber \\&
+\frac{384}{5}\zeta_2^2 +1520\zeta_3 -\frac{9008}{9} + \mathcal{O}\left(\epsilon\right) \bigg].
\end{align}
The color factors above are defined in Eqs. (185) - (188) of reference \cite{vanRitbergen:1998pn} for a $SU(N_c)$ gauge group ($T_F = 1/2$).

Our cross-checks were as follows. First of all, we observed
that the gauge-dependent parameters of our calculation dropped out of our final results.
Most of our master integrals were computed twice using different choices for the finite integrals and it was gratifying to see that we could often produce uniform weight integrals by mapping our finite integrals back to conventional ones (see {\it e.g.} Eq. (\ref{eq:B_10_20339})).
Due to the simplicity of finite Feynman integrals \cite{vonManteuffel:2017myy}, we found that it was possible to check all our non-planar master integrals to a relative precision of $10^{-4}$ numerically using {\tt FIESTA 4} \cite{Smirnov:2015mct}.
We were also able to successfully check some of our integrals analytically against the results of
\cite{Lee:2017mip,Lee:2019zop}.
Furthermore, we agreed with the higher-order pole predictions of \cite{Moch:2005id} and the known result for the $N_f^2$ part of the bare quark form factor \cite{Lee:2017mip}.

Finally, we extracted the cusp anomalous dimensions from the $\epsilon^{-2}$ poles:
\begin{align}
\label{eq:cuspquark}
&\Gamma_4^q\Big|_{N_f^2} = \txblu{C_A C_F} \bigg[-\frac{224}{15}\zeta_2^2+ \frac{2240}{27}\zeta_3 - \frac{608}{81} \zeta_2 +\frac{923}{81}\bigg]
\nonumber \\&
\qquad \qquad+\txblu{C_F^2} \bigg[\frac{64}{5}\zeta_2^2 -\frac{640}{9}\zeta_3 + \frac{2392}{81}\bigg],
\\&
\Gamma_4^q\Big|_{N_{q \gamma}N_f} = 0,
\\
\label{eq:cuspgluon}
&\Gamma_4^g\Big|_{N_f^2} = \txblu{C_A^2} \bigg[-\frac{224}{15}\zeta_2^2+ \frac{2240}{27}\zeta_3 - \frac{608}{81} \zeta_2 +\frac{923}{81}\bigg]
\nonumber \\&
\qquad \qquad +\txblu{C_A C_F} \bigg[\frac{64}{5}\zeta_2^2 -\frac{640}{9}\zeta_3 + \frac{2392}{81}\bigg].
\end{align}
As predicted by the Wilson loop picture, the color structure $C_F^2 N_f^2$ in the gluon form factor enters the cusp with zero coefficient, as does the quartic Casimir term. In fact, we have confirmed by direct calculation that Casimir scaling holds for the $N_f^2$ contributions.
Eq. (\ref{eq:cuspquark}) is in agreement with the analytic result of reference \cite{Ruijl:2016pkm} and Eq. (\ref{eq:cuspgluon}) represents the first direct extraction of $\Gamma_4^g|_{N_f^2}$.
\vspace*{-1ex}

\section{Summary}
\vspace*{-1ex}
In this paper, we calculated previously-unknown contributions to four-loop quark and gluon form factors with two closed fermion loops.
We also derived the corresponding contributions to the cusp anomalous dimensions. 
The master integrals were computed by direct integration of finite
integrals in the Feynman parametric representation.
For two integral topologies, we had to perform simple changes of
variables to render the representation linearly reducible and thus accessible
to an integration with {\tt HyperInt}.

%\vspace{.59 in}

\vspace{2ex}
\noindent {\em Acknowledgments:}
We would like to thank Roman Lee for helpful discussions of parametric
annihilators.
We gratefully acknowledge Erik Panzer for helpful discussions, in particular
for pointing out higher order parametric annihilators to us, for ongoing collaborations related to this work, and for feedback on the manuscript.
We are indebted to Hubert Spiesberger for essential help and the PRISMA excellence cluster for generous financial support with computing resources.
Our computations were carried out in part on the supercomputer Mogon at Johannes Gutenberg University Mainz (\url{www.hpc.uni-mainz.de}), 
and we wish to thank the Mogon team for their technical support.
We thank Dalibor Djukanovic for generously providing additional
computing resources at the Helmholtz Institute at Johannes Gutenberg University
Mainz.
This work employed computing resources provided by the
High Performance Computing Center at Michigan State University,
and we gratefully acknowledge the HPCC team for their help and support.
This work was supported in part by the National Science Foundation under Grant No. 1719863.
RMS was supported in part by the European Research Council through grant 647356 (CutLoops). Our figures were generated using {\tt Jaxodraw} \cite{Binosi:2003yf}, based on {\tt AxoDraw} \cite{Vermaseren:1994je}.

\bibliography{ff4lnf2}

\end{document}